\documentclass[fleqn,usenatbib]{mnras}
\usepackage[T1]{fontenc}
\usepackage{newtxtext,newtxmath}
\usepackage{graphicx}
\usepackage{amsmath}
\usepackage{booktabs}
\usepackage{xcolor}
\usepackage{hyperref}
\usepackage{lineno}

\newcommand{\beq}{\begin{eqnarray}}
\newcommand{\eeq}{\end{eqnarray}}

\begin{document}
\label{firstpage}
\pagerange{\pageref{firstpage}--\pageref{lastpage}}
	
\title[Recurrent double-peaked $\gamma$-ray sub-flares]{Recurrent double-peaked $\gamma$-ray sub-flares in PKS 1424--418}
	
\author[N. Wang et al.]{
	Na Wang$^{1}$,
	Tong Liu$^{1}$\thanks{E-mail: tongliu@xmu.edu.cn}
	and Jingran Xu$^{2}$ \\
	$^{1}$Department of Astronomy, Xiamen University, Xiamen, Fujian 361005, China\\
	$^{2}$CAS Key Laboratory of Space Astronomy and Technology, National Astronomical Observatories, Chinese Academy of Sciences, Beijing 100101, China
	}
	
\date{Accepted XXX. Received YYY; in original form ZZZ}
	
\pubyear{\the\year}
	
\maketitle
	
\begin{abstract}
We report a hint of recurrent double-peaked $\gamma$-ray sub-flares in PKS~1424--418 during two similar major active epochs, MJD 56117--56498 and MJD 59669--59978. The candidate double-peaked sub-flares show a characteristic intra-subflare peak separation of $\sim 11.8$ d and a neighbouring-subflare spacing of $\sim 67$ d. Red-noise Monte Carlo tests indicate that the candidate recurrent double-peaked morphology is not easily reproduced by stochastic variability alone, with false-alarm probabilities of $p_{\rm a}=0.0216$ and $p_{\rm b}=0.0006$ for the two active epochs, respectively. In the simulations of the entire \textit{Fermi} Large Area Telescope (LAT) light curve, no red-noise realisation reproduces both epoch-like structures. The spectral energy distribution (SED) modelling shows similar radiative properties for the two peaks of one double-peaked candidate, while changes in the Doppler factor appear to play an important role in the flaring activity. Together with the previously reported near-zero lag between the millimetre and $\gamma$-ray emission, these results suggest that the recurrent activity may be associated with the compact mm/sub-mm core region. We discuss a possible structured-jet scenario in which a single disturbance propagates through a chain of quasi-stationary recollimation shocks. In this picture, repeated interactions between the disturbance and the shock structure may qualitatively account for both the double-peaked sub-flare profiles and the observed $\sim 11.8$ d and $\sim 67$ d time-scales. Further active epochs are required to determine whether this similarity represents a recurrent long-term behaviour.
\end{abstract}
	

\section{Introduction}\label{intro}
	
An active galactic nucleus (AGN) hosts a supermassive black hole (SMBH) at its centre, which accretes surrounding material via an accretion disc and launches relativistic jets. Blazars are AGNs with relativistic jets oriented close to our line of sight, leading to high luminosity, strong polarisation, and rapid large-amplitude variability \citep{1995PASP..107..803U}. The broadband spectral energy distribution (SED) of blazars exhibits a characteristic double-humped structure \citep[e.g.,][]{1998MNRAS.299..433F}. The low-energy hump arises from synchrotron radiation (Syn) emitted by relativistic electrons spiralling in the jet magnetic field. The high-energy hump is produced by inverse Compton scattering (IC) of low-energy seed photons \citep[e.g.,][]{2010ApJ...716...30A}. If the seed photons originate from the Syn emission itself, the process is referred to as synchrotron self-Compton \citep[SSC, e.g.,][]{1974ApJ...188..353J}. If the seed photons instead come from external radiation fields such as the accretion disc, the broad-line region (BLR), or the dusty torus (DT), the mechanism is classified as external Compton (EC). This scenario often produces a strong Compton dominance in the high-energy component. 
	
Based on emission-line strengths, blazars are subdivided into flat-spectrum radio quasars (FSRQs), which exhibit strong broad lines and are typically EC-dominated in their high-energy emission, and BL Lac objects (BL Lacs), which have weak or absent lines. Furthermore, blazars are categorised by their Syn peak frequency ($\nu_{s}$) into high-Syn-peaked (HSP; $\nu_{s} > 10^{15}\,$Hz), intermediate-Syn-peaked (ISP; $10^{14}<\nu_{s}<10^{15}\,$Hz), and low-Syn-peaked (LSP; $\nu_{s}<10^{14}\,$Hz) \citep[e.g.,][]{1995ApJ...444..567P}.
	
\begin{figure*}
	\centering
	\includegraphics[width=0.8\textwidth]{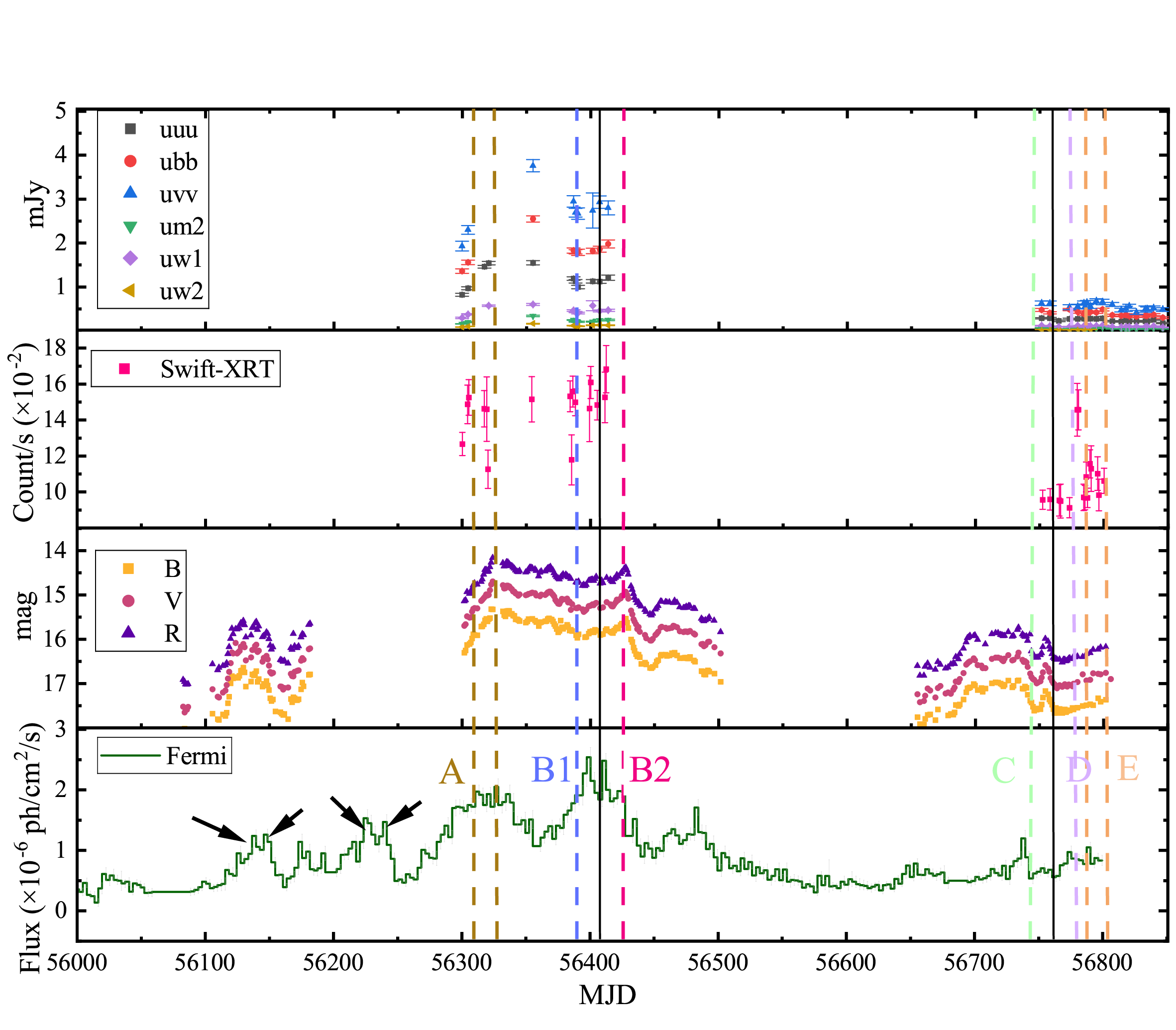}
	\caption{Multiwavelength light curves of PKS 1424--418. From top to bottom: the first panel displays \textit{Swift} Ultraviolet/Optical Telescope (UVOT) data (UVW1, UVM2, UVW2, V, B, and U filters); the second panel displays \textit{Swift} X-ray Telescope (XRT) data; the third panel displays optical B-band (yellow points), V-band (pink points), and R-band (violet points) data from SMARTS; and the fourth panel displays the \textit{Fermi} Large Area Telescope (LAT) light curve. Vertical lines denote the flaring and quiescent intervals considered in this work: flaring states A, B1, and B2 and quiescent states C, D, and E.}
	\label{fig1}
\end{figure*}

PKS 1424--418 is classified as an FSRQ in the fourth \textit{Fermi} Large Area Telescope (LAT) source catalogue \citep[4FGL J1427.9$-$4206;][]{2023ApJS..265..31} at a redshift of $z=1.522$. It is located at $\mathrm{RA}=216.987^\circ$ and $\mathrm{Dec}=-42.106^\circ$, and hosts an SMBH with a mass of $\sim 4.5\times10^9 M_\odot$ \citep{2004ApJ...602..103F}. The broadband variability of this source has been studied extensively. \citet{2014A&A...569A..40B} modelled several flaring-state SEDs with a one-zone leptonic model including external seed photons from the BLR and DT, and found that the variability can be explained by changes in the jet particle flux and energy distribution. More recently, \citet{2021MNRAS.501.2504A} identified two major flaring episodes and reported tight correlations among the $\gamma$-ray, optical, and infrared bands, with SED modelling suggesting that the $\gamma$-ray emission region lies beyond the BLR. A later $\gamma$-ray spectral study also suggested that the dissipation site in PKS 1424--418 may vary with the source activity state \citep{2024ApJ...968L...1A}. These studies mainly focused on broadband variability, spectral properties, and SED modelling of major flaring episodes, while recurrent double-peaked substructures in the $\gamma$-ray light curve have not yet been systematically examined.

In addition to these large-scale flaring events, the $\gamma$-ray light curve of PKS 1424--418 also exhibits a possible recurrent double-peaked structure. Complex substructures during blazar flares can in principle arise within one-zone or multi-zone models, for example through time-dependent particle injection, Doppler-factor variations, or changing contributions from multiple emitting components \citep[e.g.,][]{1997A&A...320...19M, 2013MNRAS.431..824B,2019ApJ...883..137P,2020ApJ...902...61L}. Magnetic reconnection may also produce rapid substructures in blazar jets \citep{2020NatCo..11.4176S}. A related ``peak-in-peak'' variability pattern was reported in the blazar 3C 279 by \citet{2020NatCo..11.4176S}, who interpreted minute-scale flares and hour-scale peak separations in terms of relativistic magnetic reconnection. The much longer characteristic intervals observed in PKS 1424--418 indicate that the applicability of such an interpretation needs to be examined separately. We note that three sub-flares of PKS 1424--418 during Modified Julian Date (MJD) 56117--56498 show candidate double-peaked profiles, and that comparable profiles are also present during MJD 59669--59978 (Figure \ref{fig2}). The repeated appearance of this pattern motivates us to test whether such structures can be distinguished from stochastic red-noise variability. The key question is therefore what statistical support exists for the candidate recurrent double-peaked morphology and what physical process, if any, may produce the observed recurrent timing pattern.

The central objective of this work is to examine the statistical support for the candidate recurrent double-peaked $\gamma$-ray sub-flares in PKS 1424--418 and to explore their possible physical origin. We combine multiwavelength light curves, quasi-simultaneous broadband SED modelling, and \textit{Fermi}-LAT $\gamma$-ray timing analysis. The SED modelling is based on six quasi-simultaneous broadband data sets selected within MJD 56000--56800, including the two peaks of one double-peaked candidate and comparison flaring and quiescent states, and is used to examine whether the double-peaked structure is accompanied by distinct changes in the radiative properties. The timing analysis focuses on two major active epochs, MJD 56117--56498 and MJD 59669--59978, where the hint of recurrent double-peaked morphology is examined and tested against a red-noise null hypothesis. Based on these results, we further discuss possible physical mechanisms that could account for both the intra-subflare peak separation and the neighbouring-subflare spacing. The paper is organised as follows. Section~\ref{data} describes the multiwavelength observations, data-reduction procedures, and the timing-analysis pipeline used to identify and statistically test the double-peaked sub-flares. Section~\ref{result} presents the results and discussion, including the SED modelling, timing and statistical analysis of the candidate recurrent double-peaked morphology, possible physical interpretations, and a variability comparison between the two major active epochs. Section~\ref{Conclusion} summarises our main conclusions.

\section{Multiwavelength data}
\label{data}
	
In this section, we compile the multiwavelength light curves employed in the analysis of PKS 1424--418, detailing in turn the extraction and reduction methods for \textit{Swift} ultraviolet/optical telescope (UVOT), \textit{Swift} X-ray telescope (XRT), Small and Moderate Aperture Research Telescope System (SMARTS, optical B-, V-, and R-bands) and \textit{Fermi}-LAT ($\gamma$-ray).
	
\subsection{\textit{Swift}-UVOT}

The Neil Gehrels \textit{Swift} Observatory is a multiwavelength space observatory launched by the National Aeronautics and Space Administration (NASA) in 2004, equipped with the X-ray Telescope (XRT; \citealt{2005SSRv..120..165B}), the Ultraviolet/Optical Telescope (UVOT; \citealt{2005SSRv..120...95R}), and the Burst Alert Telescope (BAT; \citealt{2005SSRv..120..143B}). Operating over 0.2--150\,keV, \textit{Swift} provides rapid localisation and follow-up of transient high-energy phenomena, making it a cornerstone instrument for studies of $\gamma$-ray bursts, AGN, and other explosive cosmic events. The \textit{Swift}-UVOT provides imaging in six filters, namely V (546.8 nm), B (439.2 nm), U (346.5 nm), UVW1 (260.0 nm), UVW2 (192.8 nm), and UVM2 (224.6 nm), covering 180--600\,nm. All UVOT data within the interval MJD 56000--56800 were processed using the standard HEASoft 6.26.1 threads\footnote{\url{https://www.swift.ac.uk/analysis/uvot/index.php}}. Raw images in each filter were co-added using \textit{UVOTIMSUM}, and source and background magnitudes were extracted with \textit{UVOTSOURCE} using 5\arcsec\ and 20\arcsec\ apertures, respectively. All magnitudes were corrected for Galactic extinction assuming $E(B-V)=0.093$\,mag and adopting the filter-dependent extinction coefficients from \citet{2015MNRAS.454..353R}.
	
\subsection{\textit{Swift}-XRT}
	
The Neil Gehrels \textit{Swift} Observatory hosts the \textit{Swift}-XRT, which operates in photon counting (PC) and windowed timing (WT) modes over 0.2--10\,keV, enabling photometry with $\lesssim$10\,ms resolution \citep{2004ApJ...611.1005G}. To assemble X-ray data quasi-simultaneous with the $\gamma$-ray observations of PKS 1424--418, we retrieved data for MJD 56000--56800 from the High Energy Astrophysics Science Archive Research Center (HEASARC) archive\footnote{\url{https://heasarc.gsfc.nasa.gov/db-perl/W3Browse/w3browse.pl}}. These data were processed following the standard threads\footnote{\url{https://www.swift.ac.uk/analysis/uvot/index.php}} for Level I data. Standard processing with \textit{xrtpipeline} (HEASoft 6.26.1) produced good time intervals (GTIs). For the PC data, the source counts were then extracted with \textit{xselect} using a circular region with a radius of 20 pixels (47\arcsec), while the background was extracted from an annular region with inner and outer radii of 51 pixels (120\arcsec) and 85 pixels (200\arcsec), respectively. Light curves and spectra were extracted from the Level-II WT and PC mode data using \textit{xselect}. We generated auxiliary response files via \textit{xrtmkarf}, using the swxwt0to2s6\_20131212v015.rmf (WT) and swxpc0to12s6\_20130101v014.rmf (PC) response matrices, and binned the spectra with \textit{grppha} to $\ge20$ (WT) or $\ge10$ (PC) counts per bin. Finally, the spectra were fitted in \textit{xspec} \textsl{12.10.1} with a redshifted power-law model \citep{1996ASPC..101...17A}.
	
\subsection{SMARTS}

The SMARTS at the Cerro Tololo Inter-American Observatory (CTIO) conducts Johnson B, Johnson V, and Cousins R photometric monitoring of all blazars detected by the \textit{Fermi}-LAT that are visible from Chile. Given the nearly identical variability across these filters \citep{2012ApJ...756...13B}, we adopt the Johnson B light curve as a proxy for the combined optical (B-, V-, and R-bands). We extracted SMARTS B-, V-, and R-band observations spanning MJD 56000--56800 from the website\footnote{\url{http://www.astro.yale.edu/smarts/glast/home.php}}.
	
\begin{figure*}
	\centering
	\includegraphics[width=0.8\textwidth]{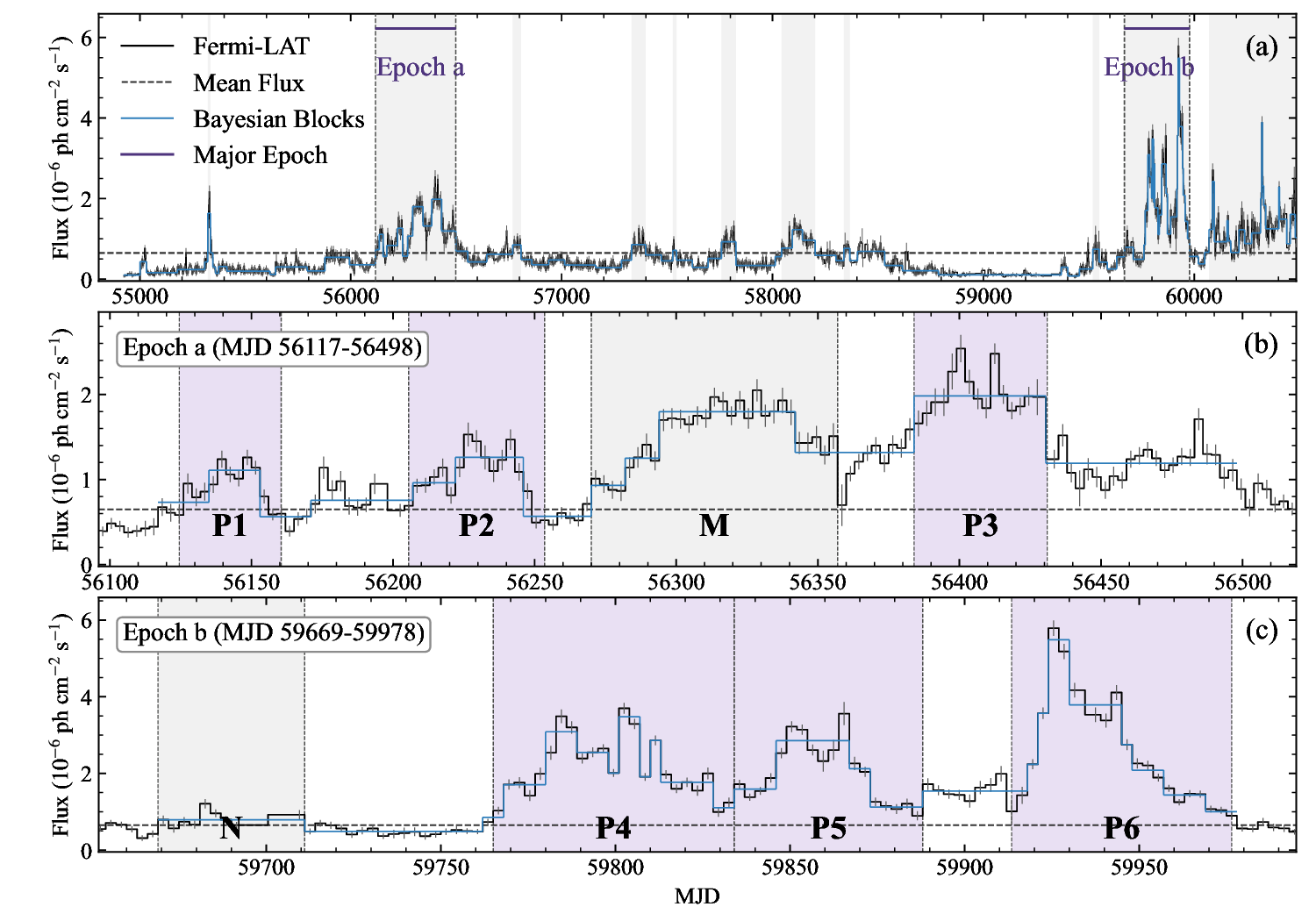}
	\caption{Bayesian-block identification of the major $\gamma$-ray flaring epochs and candidate sub-flares of PKS 1424--418. Panel (a) shows the full \textit{Fermi}-LAT $\gamma$-ray light curve of PKS 1424--418 together with the Bayesian-block representation. The light curve is segmented into flaring episodes identified by HOP grouping, with grey regions indicating high states. Panels (b) and (c) present zoomed-in views of Epoch a and Epoch b, respectively.}
	\label{fig2}
\end{figure*}	
	
\subsection{\textit{Fermi}-LAT}
	
The \textit{Fermi}-LAT is a pair-conversion $\gamma$-ray detector aboard the \textit{Fermi} Gamma-ray Space Telescope, launched in 2008. It is sensitive to photons from $\sim$20\,MeV up to several hundred GeV, and into the TeV regime, with an energy resolution of 15 per cent at 100\,MeV, improving at higher energies \citep{2009ApJ...697.1071A}. With a field of view of $\sim2.4\,$sr and operating in an all-sky scanning mode every $\sim3\,$h, \textit{Fermi}-LAT continuously monitors $\gamma$-ray sources, yielding high-cadence light curves and enabling SED analysis. We processed the \textit{Fermi}-LAT Pass 8 data in the 0.1--300\,GeV range from MJD 56000 to 56800 for PKS 1424--418. We used \textit{Fermitools} to uniformly preprocess the irregular and discontinuous \textit{Fermi}-LAT observational data. We took the \textit{Fermi}-LAT photon and spacecraft FITS files (*.fits) and converted them into comma-separated-value (CSV) format (*.csv), so they could be used in the Likelihood Analysis With Python website\footnote{\url{https://fermi.gsfc.nasa.gov/ssc/data/analysis/scitools/python_tutorial.html}}. First, we applied \textit{gtselect} to extract 0.1--300\,GeV events within a 10$^\circ$ region of interest (ROI) centred on the source. We also imposed a zenith angle cut ($<90^\circ$) to suppress Earth limb contamination. GTIs were established with \textit{gtmktime}, and counts maps were generated with \textit{gtbin}; exposure cubes and maps were then produced with \textit{gtltcube} and \textit{gtexpmap}, respectively \citep{2020ApJS..247...33A}. A comprehensive source model was built by combining the Galactic interstellar emission model (gll\_iem\_v07.fits), the isotropic background\footnote{\url{https://fermi.gsfc.nasa.gov/ssc/data/access/lat/BackgroundModels.html}} (iso\_P8R3\_SOURCE\_V3\_v1.txt), and all 4FGL point sources within 15$^\circ$ via the \textsl{4FGL}\textit{xml.py} script. The diffuse components were produced with \textit{gtdiffresp}. Finally, a binned maximum likelihood fit was performed using the \textit{gtlike} task, with source significance quantified by the test statistic (TS), $\mathrm{TS}=2\,\Delta\ln\mathcal{L}$, and the resulting photon spectrum was extracted using \textit{Fermipy} \citep{2017ICRC...301W}. The 3-day-binned $\gamma$-ray light curve was constructed using only bins with $\mathrm{TS} > 25$, corresponding to a detection significance of approximately $5\sigma$. All data points shown in Figure \ref{fig2} satisfy this criterion.
	
\subsection{Double-peaked Identification and Statistical Validation}
\label{double-peak}
	
This section outlines the procedure for identifying candidate double-peaked $\gamma$-ray sub-flares and for testing their statistical significance against stochastic red-noise variability. The results are summarised in Section \ref{Timing properties}. The timing analysis presented below is based on the \textit{Fermi}-LAT light curve, which provides continuous and homogeneous coverage across the two major active epochs considered in this work. In contrast, the lower-energy light curves are used mainly for constructing quasi-simultaneous SEDs and for supporting the physical interpretation, because of their sparse and highly non-uniform sampling.
	
\subsubsection{Window identification and model comparison}
\label{sec:window_model}

To define the major activity epochs and candidate sub-flare windows, we applied the Bayesian Blocks (BB) algorithm and HOP algorithm to the \textit{Fermi}-LAT light curve \citep{1998ApJ...498..137E,2013ApJ...764..167S,Wagner2021}. First, we applied the BB algorithm to represent the light curve as a piecewise-constant function. Each change point between adjacent blocks indicates a flux variation significant at the 3$\sigma$ level relative to the previous block. The blocks higher than their neighbouring blocks were then identified as peaks, and decreasing blocks were traced sequentially from each peak in both directions until the flux dropped below the mean-flux threshold. The continuously connected active blocks were then grouped into high-state HOP groups \citep{2025MNRAS.537..332A}. Neighbouring HOP groups separated by less than 60 d were further merged into broader active complexes. The 60 d scale was used only to define the broad active epochs and was not used in the subsequent double-peaked classification. The two active complexes containing the major outbursts were selected as Epoch a and Epoch b, corresponding to MJD 56117--56498 and MJD 59669--59978, respectively.

Candidate sub-flare windows were then determined locally within each selected active epoch. Specifically, the BB algorithm and HOP algorithm were repeated within each epoch, and blocks that exceeded the activity threshold and were local maxima relative to their neighbouring blocks were identified as local peak blocks. The separations between adjacent local peak blocks showed a clear gap: short separations clustered at 9--18 d, whereas longer separations were $\geq54$ d. To avoid overly fragmented local BB/HOP segmentation, especially in Epoch b where neighbouring high-state peaks could be over-separated into multiple independent events, we adopted the midpoint of this gap, 36 d, as an operational merging threshold for defining local candidate windows. Adjacent peak blocks separated by $\leq36$ d were assigned to the same candidate sub-flare, whereas peak blocks separated by larger intervals were treated as distinct candidate events. This threshold was used only for window definition and not as a statistical criterion for identifying double-peaked events.

After the candidate cores were defined, the final sub-flare boundaries were refined by extending the candidate regions outward until the lower flux-error envelope approached the activity threshold. When neighbouring windows overlapped after this boundary refinement, the original BB/HOP-defined split boundary was retained to avoid mixing adjacent candidate events. This procedure yielded four candidate windows in Epoch a, labelled P1, P2, M, and P3, and four candidate windows in Epoch b, labelled N, P4, P5, and P6. These windows are shown in Figure \ref{fig2}. The candidate windows with sufficient data are listed in Table \ref{tab1}, and their subsequent classification is summarised in Section \ref{Timing properties}.
	
For each final candidate window, we fitted the \textit{Fermi}-LAT data with both a single-Gaussian (1G) and a double-Gaussian (2G) model. Assuming Gaussian errors, we computed the Akaike Information Criterion (AIC) and Bayesian Information Criterion (BIC) for both the 1G and 2G fits \citep{1974ITAC...19..716A,1978AnSta...6..461S}. The relative preference for the 2G description was quantified using
$\Delta\chi^2=\chi^2_{\rm 1G}-\chi^2_{\rm 2G}$, $
\Delta{\rm AIC}={\rm AIC}_{\rm 1G}-{\rm AIC}_{\rm 2G}$, and $\Delta{\rm BIC}={\rm BIC}_{\rm 1G}-{\rm BIC}_{\rm 2G}$,
where positive values favour the 2G model. A candidate was classified as a fit-supported double-peaked candidate when the information criteria strongly favoured the 2G model, for which we adopted conservative thresholds of $\Delta{\rm AIC}\geq10$ and $\Delta{\rm BIC}\geq10$ \citep{2002msma.book.....B,1995JASA...90..773K}. Candidates satisfying only weaker information-criterion improvements, e.g. $\Delta{\rm AIC}>5$ and $\Delta{\rm BIC}>5$, were treated as weaker double-peaks.

\begin{table}
	\centering
	\caption{Comparison of single-Gaussian (1G, blue dashed) and double-Gaussian (2G, black solid) models for the candidate sub-flare windows. Column (2) gives the candidate MJD interval, Column (3) gives the fitted intra-subflare peak separation, Column (4) gives the spacing to the next sub-flare, and Columns (5)--(7) give the model-comparison statistics.}
	\label{tab1}
	\footnotesize
	\setlength{\tabcolsep}{2pt}
	\begin{tabular}{@{}lcccccc@{}}
		\toprule
		Sub-flare & MJD & $\Delta t_{\rm peak}$ (d) & $\Delta t_{\rm next} $ (d) 
		& $\Delta\chi^2$ & $\Delta{\rm AIC}$ & $\Delta{\rm BIC}$ \\
		
		(1) & (2) & (3) & (4) & (5) & (6) & (7) \\
		\midrule
		P1 & 56124--56160 & 10.02  & 78  & 14.94  & 8.94   & 7.24 \\
		P2 & 56205--56253 & 15.51 & 102 & 22.15  & 16.15  & 13.66 \\
		M  & 56269--56358 & 25.31 & 72  & 5.94   & -0.06  & -4.17 \\
		P3 & 56384--56431 & 13.58 & --  & 22.55  & 16.55  & 14.43 \\
		P4 & 59764--59834 & 7.44  & 63  & 103.33 & 97.33  & 93.93 \\
		P5 & 59834--59913 & 13.41 & 60  & 61.56  & 55.56  & 51.79 \\
		P6 & 59913--59976 & 11.01 & --  & 187.23 & 181.23 & 178.09 \\
		\bottomrule
	\end{tabular}
	
	\raggedright
	\footnotesize
\end{table}	

The results of 1G and 2G model fits are shown in Table \ref{tab1} and Section \ref{Timing properties}. The AIC and BIC comparison quantifies only the local preference for a 2G description within each candidate window, whereas the false-alarm probability associated with the hint of recurrent double-peaked structure relative to the red-noise null hypothesis is evaluated separately using the Monte Carlo procedure described in Section~\ref{sec:red_noise_mc}.

\subsubsection{Timing analysis}
\label{timing}

Before evaluating the statistical significance of candidate recurrent double-peaked structure, we first estimated the power spectral density (PSD) slopes to define the red-noise null model. We then applied the autocorrelation function (ACF), partial autocorrelation function (PACF), and double power spectrum (DPS) to search for possible characteristic lag scales and to compare them with red-noise expectations. These diagnostics are used only as auxiliary timing checks and are not interpreted as independent detections of a unique time delay or strict periodicity.

\textit{Power spectral density.}
Blazar $\gamma$-ray light curves are commonly dominated by stochastic red-noise variability, whose PSDs are often described by a power-law model, $P(\nu)\propto \nu^{-k}$, where $\nu$ is the Fourier frequency and $k$ is the PSD slope \citep[e.g.,][]{2010ApJ...722..520A,2013ApJ...773..177N}. We estimated the power-law variability index $k$ using the power spectral response (PSRESP) method \citep{2002MNRAS.332..231U,2014MNRAS.445..428M,Wang2026}. For each trial value of $k$, we generated $10^3$ simulated light curves with the same temporal sampling and duration as the observed \textit{Fermi}-LAT light curves. The simulations were constructed to match both the assumed PSD shape and the observed flux distribution \citep{2013MNRAS.433..907E,2015arXiv150306676C}. The simulated light curves were analysed in the same way as the observed data when estimating the PSD, thereby accounting for red-noise leakage and aliasing effects.
The best-matching input PSD slopes are $k_{\rm a}=1.90^{+0.35}_{-0.15}$ for Epoch a, $k_{\rm b}=1.35^{+0.25}_{-0.22}$ for Epoch b, and $k_{\rm full}=1.48^{+0.03}_{-0.03}$ for the entire light curve. These slopes were then adopted as the red-noise null models for the timing analysis and Monte Carlo tests.

\textit{Autocorrelation function.}
The ACF quantifies the linear correlation between a time series and its time-shifted version. For light curves containing repeated temporal structures, such as a delayed component superposed on the intrinsic signal or recurrent flare profiles with a characteristic separation, the ACF may exhibit a local maximum at the corresponding lag \citep{2011A&A...528L...3B,2015ApJ...809..100B}. In red-noise-dominated variability, however, stochastic low-frequency fluctuations can also produce apparent ACF peaks. Therefore, individual ACF features are not interpreted in isolation. We computed the ACF for each epoch and compared the observed values with red-noise confidence levels derived from simulated light curves generated using the PSRESP-derived PSD slope of the corresponding epoch.

\textit{Partial autocorrelation function.}
The PACF provides a complementary analysis by quantifying the residual correlation at each lag after removing the linear dependence on all shorter lags \citep[e.g.,][]{2002itsa.book.....B}. Features appearing in the ACF at a given lag may simply be inherited through the correlations propagated from shorter lags, whereas the PACF helps isolate the direct contribution at that lag. This distinction is particularly useful in the red-noise regime, where low-frequency power leakage can cause the ACF to show broad, correlated excesses over multiple lags. We computed the PACF for the same epoch light curves and assessed its features using the same ensemble of red-noise simulations as that adopted for the ACF.

\textit{Double power spectrum.}
The DPS, originally introduced by \citet{2011A&A...528L...3B}, provides a complementary Fourier-domain analysis to the ACF and PACF. It is based on the fact that a delayed or repeated component in a light curve introduces a periodic modulation into its first power spectrum (FPS). For a delayed component with relative amplitude $a$ and time delay $t_0$, the FPS can be written as $|\tilde{S}(f)|^2=|\tilde{s}(f)|^2\left[1+a^2+2a\cos(2\pi f t_0)\right]$, where $\tilde{S}(f)$ and $\tilde{s}(f)$ are the Fourier transforms of the observed and intrinsic light curves, respectively, and $f$ is the Fourier frequency. The modulation period in Fourier-frequency space is inversely proportional to $t_0$. Therefore, taking the power spectrum of the FPS, i.e. the DPS, can reveal the corresponding lag as a peak in the time-delay domain \citep{2011A&A...528L...3B,2015ApJ...809..100B}.

The red-noise confidence levels in the ACF, PACF, and DPS were evaluated using the same ensemble of red-noise simulations. For each epoch, we generated simulated light curves following the method of \citet{2013MNRAS.433..907E}, using the PSRESP-derived PSD slope and the same temporal sampling and flux-uncertainty pattern as the observed \textit{Fermi}-LAT light curve. The simulated light curves were processed with the same ACF, PACF, and DPS pipelines as the observed data. For each lag, we constructed the empirical distribution of the corresponding statistic from the simulation ensemble and derived the $2\sigma$ and $3\sigma$ confidence levels. The results of the timing analysis are discussed in Section \ref{Timing properties}.

\subsubsection{Red-noise Monte Carlo}
\label{sec:red_noise_mc}
	
The timing diagnostics described above can evaluate individual lag scales under the red-noise null hypothesis, but they do not by themselves determine whether the observed hint of recurrent double-peaked structure can be produced by stochastic red-noise variability. We therefore performed red-noise Monte Carlo tests for both active epochs and for the full \textit{Fermi}-LAT light curve. 

We adopted the PSRESP-derived PSD slopes obtained in Section \ref{timing}. Each simulated light curve was generated with the same temporal sampling and flux-uncertainty pattern as the observed data, following the method of \citet{2013MNRAS.433..907E}. Each simulated light curve was processed with the same BB and 1G/2G model-comparison pipeline as the observed light curve. A simulated realisation was counted as Epoch-a-like or Epoch-b-like only if it satisfied the same peak-separation and information-criterion requirements as the corresponding observed epoch. For the simulations covering the entire LAT light curve, a simulated light curve was classified as a joint recurrence only when both an Epoch-a-like and an Epoch-b-like structure were identified. Thus, the Monte Carlo test was designed to evaluate the recurrence of the double-peaked morphology as a whole, rather than the significance of an isolated local peak. For each test, we generated $10^4$ simulated red-noise light curves. We emphasise that this test evaluates the probability of reproducing the hint of recurrent double-peaked structure under the adopted red-noise null model, rather than assigning an independent Gaussian-equivalent significance to each individual sub-flare.

\section{Results and Discussion}
\label{result}
\subsection{Broadband spectral energy distribution}
\label{sed}	

\begin{figure}
	\centering
	\includegraphics[width=0.8\columnwidth]{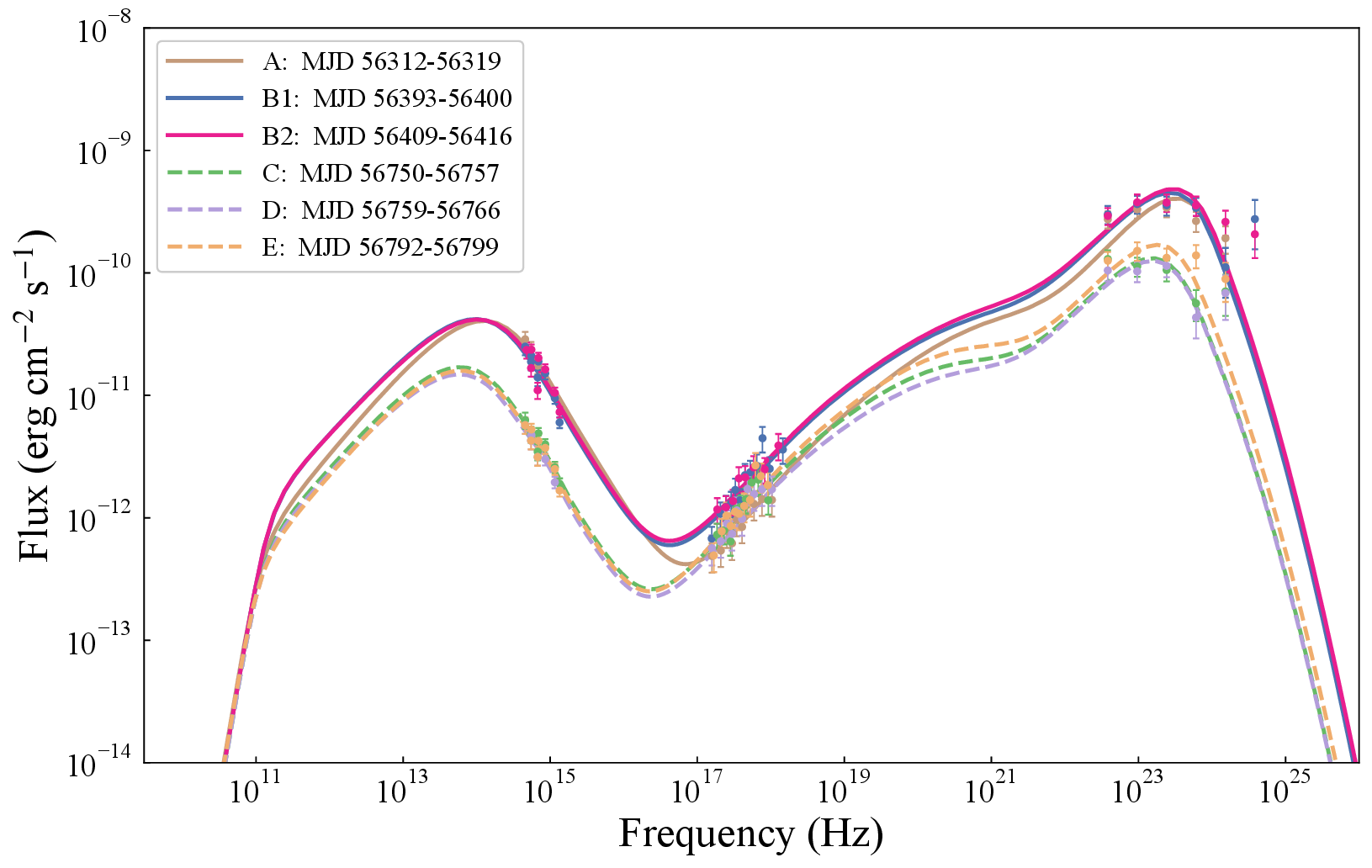}
	\caption{Overlay of the quasi-simultaneous SEDs of PKS 1424--418. Solid lines denote flaring states (A, B1, B2), and dashed lines denote quiescent states (C, D, E).}
	\label{fig3}
\end{figure}

\begin{figure*}
	\centering
	\includegraphics[width=1.5\columnwidth]{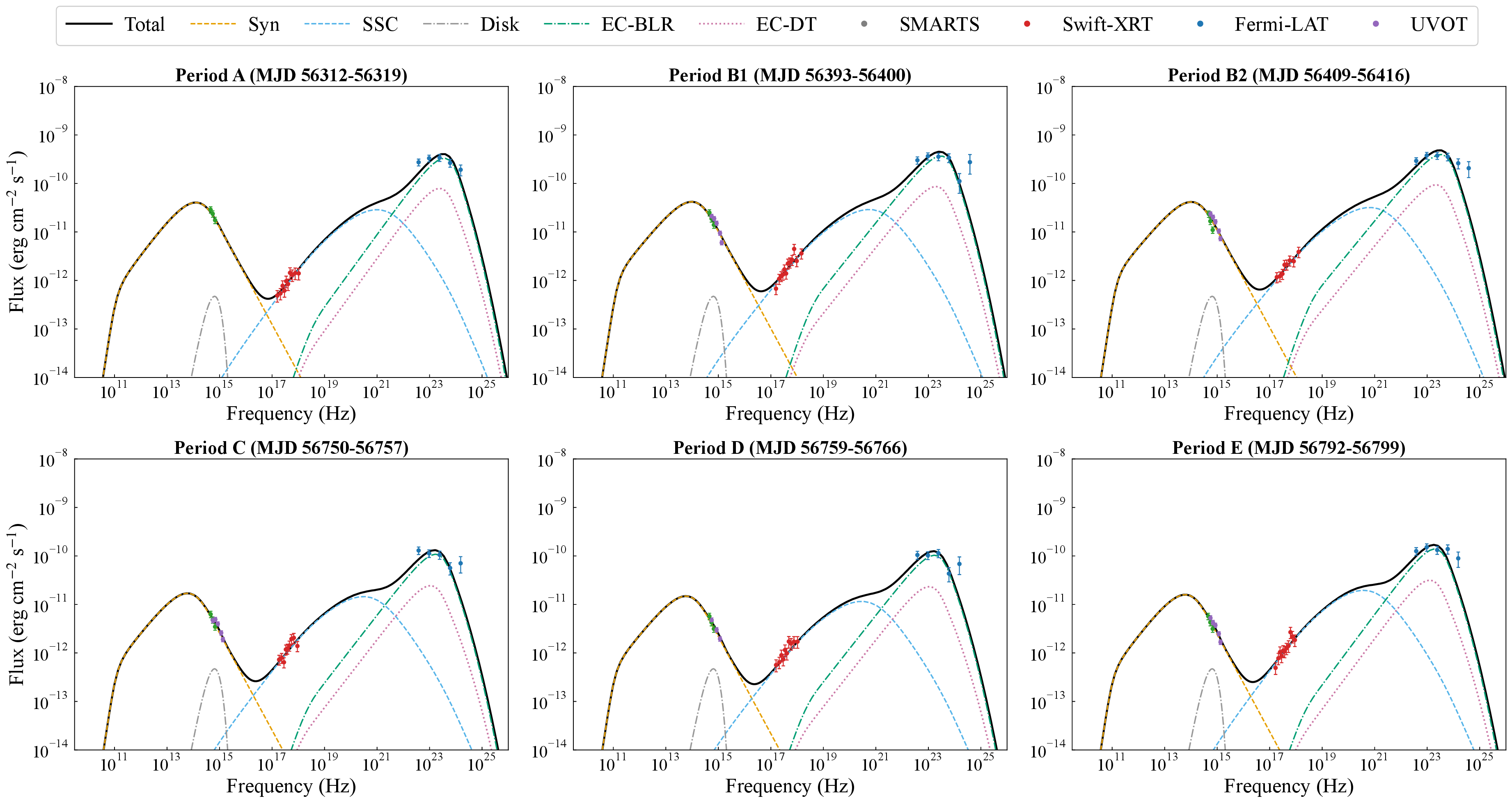}
	\caption{SEDs of PKS 1424--418 in the six selected epochs, together with the best-fitting one-zone leptonic model.}
	\label{fig4}
\end{figure*}

To investigate the broadband SEDs of PKS 1424--418 in different activity states, we selected six states with quasi-simultaneous multiwavelength coverage, namely A (MJD 56312--56319), B1 (MJD 56393--56400), B2 (MJD 56409--56416), C (MJD 56750--56757), D (MJD 56759--56766), and E (MJD 56792--56799). For each epoch, the LAT spectrum was extracted over the corresponding 7-day interval, with only spectral bins with TS $\geq 25$ retained as detections. The lower-energy data were selected from quasi-simultaneous observations within the same time window whenever available. Among them, B1 and B2 sample the two peaks of the P3 candidate, while A represents a comparison flaring state within the M window, which does not show a preference for a double-Gaussian description. C, D, and E represent relatively quiescent states. We modelled the six SEDs using the JetSeT framework within a one-zone leptonic scenario, including synchrotron emission, SSC, EC-BLR, EC-DT, and thermal emission from the accretion disc \citep{2009A&A...501..879T, 2011ApJ...739...66T, 2020ascl.soft09001T}. In the modelling, we adopted a fixed representative dissipation distance $R_{\rm H}=5.03\times10^{17} \mathrm{cm}$ for all six states, rather than fitting an independent distance for each peak. This distance is located outside the adopted BLR outer radius and well inside the dusty torus radius, so the high-energy component is mainly attributed to EC scattering of DT photons \citep{2014A&A...569A..40B,2021MNRAS.501.2504A}. In the fitting, $N$, $p$, $\gamma_{\rm break}$, $B$, and $\delta$ were treated as free parameters. The remaining model parameters were fixed to the values listed in Table~\ref{tab2}. The model reproduces the overall shape of the observed SEDs reasonably well, as shown in Figures \ref{fig3} and \ref{fig4}.

\begin{table*}
	\centering
	\caption{Model parameters for the SEDs of PKS 1424--418 during the six SED states.}
	\label{tab2}
	\setlength{\tabcolsep}{8pt}
	\begin{tabular}{llcccccc}
		\toprule
		Parameter & Symbol & A & B1 & B2 & C & D & E \\
		\midrule
		Normalisation & $N$ & 258.56 & 299.01 & 349.98 & 350.00 & 350.00 & 350.00 \\
		Low-energy index & $p$ & 1.75 & 1.64 & 1.67 & 1.59 & 1.62 & 1.54 \\
		Break Lorentz factor & $\gamma_{\rm break}$ & $2.45\times10^{3}$ & $2.19\times10^{3}$ & $2.33\times10^{3}$ & $2.00\times10^{3}$ & $2.00\times10^{3}$ & $2.20\times10^{3}$ \\
		Magnetic field (G) & $B$ & 0.63 & 0.63 & 0.59 & 0.60 & 0.58 & 0.50 \\
		Doppler factor & $\delta$ & 23.22 & 23.72 & 23.68 & 18.32 & 18.49 & 18.02 \\
		\midrule
		High-energy index & $p_{1}$ & 5.00 & 5.00 & 5.00 & 5.00 & 5.00 & 5.00 \\
		Blob radius (cm) & $R$ & $6.00\times10^{16}$ & $6.00\times10^{16}$ & $6.00\times10^{16}$ & $6.00\times10^{16}$ & $6.00\times10^{16}$ & $6.00\times10^{16}$ \\
		Distance from the SMBH (cm) & $R_{\rm H}$ & $5.03\times10^{17}$ & $5.03\times10^{17}$ & $5.03\times10^{17}$ & $5.03\times10^{17}$ & $5.03\times10^{17}$ & $5.03\times10^{17}$ \\
		Disc luminosity (erg s$^{-1}$) & $L_{\rm disc}$ & $1.00\times10^{46}$ & $1.00\times10^{46}$ & $1.00\times10^{46}$ & $1.00\times10^{46}$ & $1.00\times10^{46}$ & $1.00\times10^{46}$ \\
		BLR covering factor & $\tau_{\rm BLR}$ & 0.10 & 0.10 & 0.10 & 0.10 & 0.10 & 0.10 \\
		DT covering factor & $\tau_{\rm DT}$ & 0.08 & 0.08 & 0.08 & 0.08 & 0.08 & 0.08 \\
		DT radius (cm) & $R_{\rm DT}$ & $8.00\times10^{18}$ & $8.00\times10^{18}$ & $8.00\times10^{18}$ & $8.00\times10^{18}$ & $8.00\times10^{18}$ & $8.00\times10^{18}$ \\
		\midrule
		Magnetic energy density (erg cm$^{-3}$) & $U_B$ & $1.58\times10^{-2}$ & $1.58\times10^{-2}$ & $1.39\times10^{-2}$ & $1.43\times10^{-2}$ & $1.34\times10^{-2}$ & $9.95\times10^{-3}$ \\
		Electron energy density (erg cm$^{-3}$) & $U_e$ & $5.19\times10^{-3}$ & $9.59\times10^{-3}$ & $1.10\times10^{-2}$ & $1.66\times10^{-2}$ & $1.39\times10^{-2}$ & $2.35\times10^{-2}$ \\
		Jet power in $B$ ($10^{45}$ erg s$^{-1}$) & $P_{j,B}$ & 2.88 & 3.01 & 2.63 & 1.63 & 1.56 & 1.09 \\
		Jet power in electrons ($10^{45}$ erg s$^{-1}$) & $P_{j,e}$ & 1.03 & 2.08 & 2.08 & 1.88 & 1.61 & 2.58 \\
		Magnetisation & $\sigma_B=U_B/U_e$ & 3.04 & 1.65 & 1.26 & 0.86 & 0.96 & 0.42 \\
		\bottomrule
	\end{tabular}
	
	\vspace{2mm}
	\begin{minipage}{0.97\textwidth}
		\footnotesize
		\textit{Note.} The parameters in the upper part of the table were allowed to vary during the fitting, while those in the middle part were fixed for all epochs. The lower part lists quantities derived from the best-fitting model parameters. Additional model inputs fixed for all epochs are $z=1.522$, $\gamma_{\min}=5$, $\gamma_{\max}=5\times10^{5}$, $T_{\rm disc}=2.0\times10^{4}\,{\rm K}$, $T_{\rm DT}=10^{3}\,{\rm K}$, $R_{\rm BLR,in}=3.0\times10^{17}\,{\rm cm}$, and $R_{\rm BLR,out}=3.3\times10^{17}\,{\rm cm}$.
	\end{minipage}
\end{table*}

As shown in Table \ref{tab2}, the fitted parameters show that the flaring and relatively quiescent states are primarily distinguished by their Doppler factors $\delta$. The flaring states (A, B1, and B2) require systematically larger values of $\delta \approx 23.22$--23.72, whereas the quiescent states (C, D, and E) are characterised by lower values of $\delta \approx 18.02$--18.49. This indicates that enhanced Doppler boosting is the main driver of the high-state emission. In comparison, the magnetic field and the electron energy distribution vary only moderately across the six states. The magnetic field remains within $B \approx 0.50$--0.63 G, while the low-energy spectral index $p$ lies in the range 1.54--1.75. The break Lorentz factor $\gamma_{\rm break}$ changes from $\sim(2.0$--$2.2)\times10^3$ in the quiescent states to $\sim(2.2$--$2.5)\times10^3$ during the flaring states. These results suggest that, besides the dominant change in Doppler boosting, moderate variations in the electron energy distribution also contribute to the observed SED variability. In particular, B1 and B2, which sample the two peaks of P3, show similar radiative parameters. This suggests similar radiative conditions during the two peaks. However, state A also shows a comparably high Doppler factor, indicating that Doppler boosting alone cannot explain the hint of double-peaked structure.
	
The magnetic energy density is calculated as $U_B=\frac{B^2}{8\pi}$. The electron energy density is given by
\begin{equation}
	U_e=m_e c^2 \int_{\gamma_{\min}}^{\gamma_{\max}} \gamma\, n(\gamma)\, d\gamma,
\end{equation}
where $m_e$ is the electron mass, $\gamma$ is the Lorentz factor of the electron, and $n(\gamma)$ is the electron energy distribution. The electron energy distribution is parameterised as a broken power law \citep{2014Natur.515..376G},
\begin{equation}
	n(\gamma)=
	\begin{cases}
		N\gamma^{-p}, & \gamma_{\min}\leq \gamma \leq \gamma_{\rm break},\\[4pt]
		N\gamma_{\rm break}^{p_1-p}\gamma^{-p_1}, & \gamma_{\rm break}<\gamma\leq \gamma_{\max},
	\end{cases}
\end{equation}
where $\gamma_{\min}$ and $\gamma_{\max}$ are the minimum and maximum electron Lorentz factors, respectively, and $p_1$ is the high-energy spectral index. The jet powers carried by the magnetic field and relativistic electrons are estimated as $P_{j,B}=\pi R^2 \beta c \Gamma^2 U_B$ and $P_{j,e}=\pi R^2 \beta c \Gamma^2 U_e$, where $\beta c$ is the bulk velocity of the emitting plasma, $R$ is the radius of the emission region, and $\Gamma$ is the bulk Lorentz factor.

Using the fit parameters, we derived the corresponding energy densities and jet powers for all six states. As summarised in Table~\ref{tab2}, $U_B$ and $P_{j,B}$ generally decrease from the flaring states to the quiescent states, whereas $U_e$ and $P_{j,e}$ show an overall increasing trend. Correspondingly, the magnetisation parameter $\sigma_{B}=U_B/U_e$ decreases from 3.04 in state A to 0.42 in state E, indicating that the emitting region evolves from a more magnetically dominated state during the early flaring phase towards near-equipartition and eventually a mildly particle-dominated state in the later epochs. For a black hole mass of $M_\bullet = 4.5 \times 10^9 \, M_\odot$, the corresponding Eddington luminosity is $L_{\rm Edd} = 5.7 \times 10^{47} \, \mathrm{erg \, s^{-1}}$. The accretion disc luminosity is $\sim 1.8 \times 10^{-2} \, L_{\rm Edd}$, while the total jet power (carried by the magnetic field and relativistic electrons) is estimated at $\sim (0.56\text{--}0.89) \times 10^{-2} \, L_{\rm Edd}$. Although this jet power is lower than the accretion disc luminosity, it is comparable in order of magnitude, indicating that the jet of PKS 1424--418 represents a significant energy output channel of the system.

\subsection{Timing properties and statistical validation}
\label{Timing properties}

\begin{figure*}
	\centering
	\includegraphics[width=0.8\textwidth]{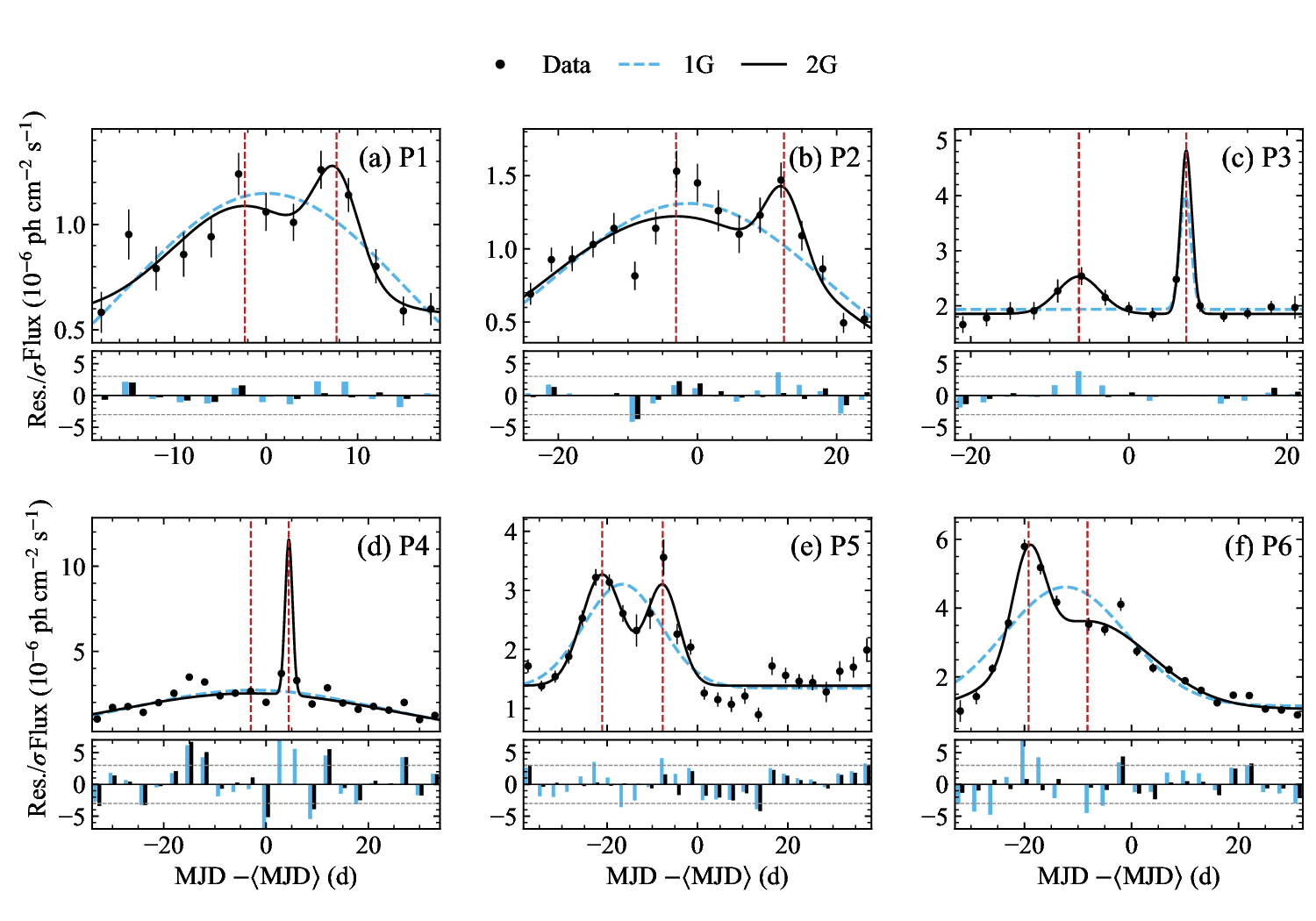}
	\caption{Single-Gaussian (1G, blue dashed) and double-Gaussian (2G, black solid) fits to the six $\gamma$-ray sub-flares (P1--P6) of PKS 1424--418.
		The red dotted curves show the two individual Gaussian components in the 2G model.
		The lower sub-panels show the residuals of the 1G and 2G fits in units of the observational uncertainty.}
	\label{fig5}
\end{figure*}

This section summarises the timing-analysis results described in Section~\ref{double-peak}, thereby defining the observational timing properties that motivate the subsequent physical interpretation. The 1G and 2G model-fitting results are shown in Figure~\ref{fig5} and summarised in Table~\ref{tab1}. According to the information-criterion thresholds adopted in this work, P2, P3, P4, P5, and P6 satisfy both $\Delta{\rm AIC}\geq10$ and $\Delta{\rm BIC}\geq10$, and are therefore classified as fit-supported double-peaked candidates. Their fitted intra-subflare peak separations lie in the range 7.4--15.5 d. P1 satisfies only the weaker information-criterion requirement and is therefore retained as a weaker double-peak. In contrast, M does not show a preference for a 2G description, while N contains too few data points for a meaningful 1G and 2G comparison. We therefore treat M and N as comparison windows and do not include them in the recurrent sequence of double-peaked candidates. Across the six candidates P1--P6, the mean separation between the two Gaussian components is $11.8 \pm2.9$ d. This time-scale characterises the typical separation between the two peaks within an individual double-peaked sub-flare. The spacing between neighbouring sub-flares is estimated from the P1--P2 sequence in Epoch a and the P4--P6 sequence in Epoch b, yielding $67 \pm10$ d. Thus, the intra-subflare peak separation of $\sim11.8$ d and the neighbouring-subflare spacing of $\sim67$ d constitute the main timing properties that any physical interpretation needs to address.

\begin{figure*}
	\centering
	\includegraphics[width=0.8\textwidth]{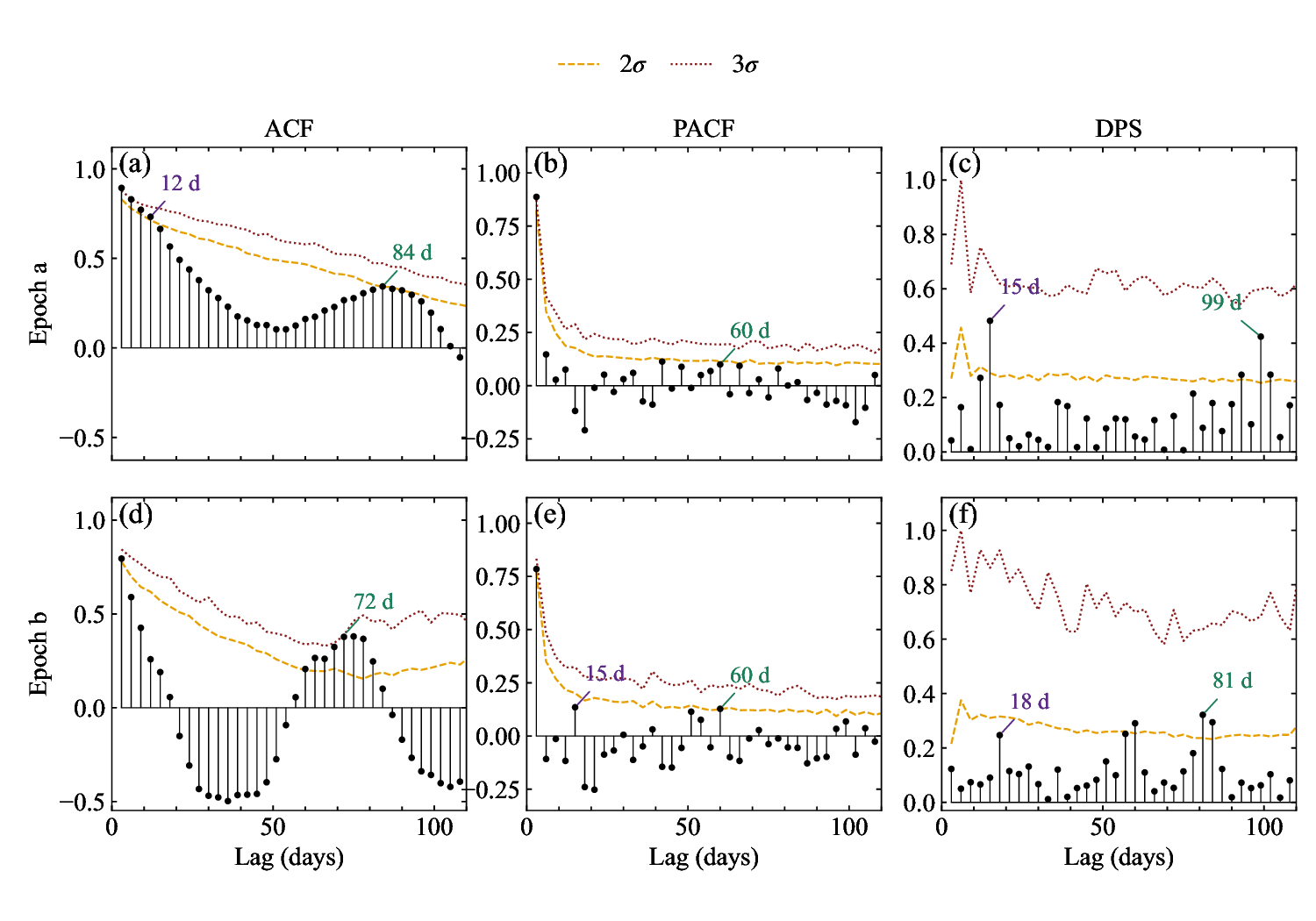}
	\caption{Timing analysis for the two major active epochs. The first and second rows correspond to Epochs a and b, respectively, while the three columns show the ACF, PACF, and normalised DPS. The dashed and dotted curves denote the red-noise $2\sigma$ and $3\sigma$ confidence levels, respectively. The marked lags are shown for comparison with the double-peaked flare morphology, but they are not interpreted as independent detections of a unique time delay or periodicity.}
	\label{fig6}
\end{figure*}

To examine whether these time-scales also appear in timing analyses that do not rely on Gaussian decomposition, we calculated the ACF, PACF, and normalised DPS for the two active epochs and compared them with red-noise simulations generated using the corresponding PSD slopes. The results are shown in Figure~\ref{fig6}. In Epoch a, the ACF, PACF, and DPS show local enhancements in the short-lag range around 12--15 d, as well as candidate features at longer lags around 60--99 d. In Epoch b, similar local features appear at comparable lag ranges, mainly around 12--18 d and 60--81 d. These candidate time-scales are broadly consistent, in order of magnitude, with the $\sim11.8$ d intra-subflare peak separation and the $\sim67 \pm 10$ d neighbouring-subflare spacing inferred from the 1G and 2G fits. However, the timing-analysis features are not statistically robust on their own. Most of the enhancements are below, or only close to, the $2$--$3\sigma$ confidence levels derived from the red-noise simulations. Therefore, these timing analyses are not used to establish the statistical significance of the double-peaked morphology. They only provide an independent consistency check that weak correlation structures occur near the candidate time-scales. The statistical significance of the candidate recurrent morphology is instead evaluated with the red-noise Monte Carlo test described below, in which the simulated light curves are processed through the same analysis pipeline as the observed data.

The red-noise Monte Carlo test provides a more direct assessment of whether the candidate recurrent double-peaked structure can be reproduced under the adopted red-noise null model. For the epoch-based simulations, 216 out of $10^4$ simulated light curves satisfy the Epoch-a-like criteria, while 6 out of $10^4$ simulated light curves satisfy the Epoch-b-like criteria. The corresponding false-alarm probabilities are $p_{\rm a}=0.0216$ and $p_{\rm b}=0.0006$. Here, ``Epoch-a-like'' and ``Epoch-b-like'' refer to simulated light curves that, after being processed with the same BB window-identification and 1G/2G model-comparison procedure as the observed data, reproduce the double-peaked morphology, neighbouring-subflare spacing, and information-criterion requirements of the corresponding observed epoch. These criteria are observation-motivated: they are based on the timing properties measured from the observed double-peak, but were kept fixed for all simulated light curves. Specifically, the simulated light curve was required to show a double-peaked sub-flare with a day-scale intra-subflare peak separation and neighbouring sub-flares with spacings similar to those observed. In the full light-curve simulations, none of the simulated light curves simultaneously reproduces both an Epoch-a-like and an Epoch-b-like structure. This probability should be interpreted as the false-alarm probability for reproducing the hint of recurrent double-peaked structure under the adopted red-noise null hypothesis, rather than as an independent significance for each individual sub-flare.

These results indicate that the individual double-peaked sub-flares should not be over-interpreted in isolation. The main feature of interest is instead the hint of recurrent double-peaked structure together with the neighbouring-subflare spacings. This overall pattern is not easily reproduced in our red-noise simulations. We therefore use this timing pattern as an observational constraint and combine it with the SED results in Section~\ref{sed} to discuss the possible physical origin of the hint of recurrent double-peaked structure.

\subsection{Candidate recurrent double-peaked sub-flares}
\label{Rep}

As shown in Table~\ref{tab1}, the candidate double-peaked structures in PKS 1424--418 have a characteristic intra-subflare peak separation of $11.8 \pm2.9$ d, while the spacing between neighbouring sub-flares is $67 \pm10$ d. Multiple GeV flares in blazars can be interpreted phenomenologically in terms of different emitting regions, magnetic reconnection, plasmoid activity, or disturbances propagating along the jet \citep[e.g.,][]{2013MNRAS.431..824B,2019ApJ...883..137P,2020ApJ...902...61L,2020NatCo..11.4176S}.

Before introducing a more specific dynamical scenario, we first consider simpler one-zone and two-zone radiative interpretations. In a one-zone model, the double-peaked structure can be phenomenologically understood as two short-lived enhancements of the same dominant emission region during its temporal evolution, for example due to two episodes of particle injection, compression, or Doppler-factor enhancement \citep[e.g.,][]{2002ApJ...581..127B,2017Natur.552..374R,2021MNRAS.501.1100R}. The quasi-simultaneous SED modelling in Section \ref{sed} shows that B1 and B2 can both be described within a similar one-zone leptonic framework. This similarity is consistent with the two peaks arising under similar radiative conditions, although the SED modelling alone cannot determine their dynamical origin. Another phenomenological interpretation is a two-zone or two-component model, in which two physically distinct, and possibly spatially separated, emission regions dominate the two peaks of the double-peaked flare \citep[e.g.,][]{2005A&A...432..401G,2008MNRAS.385L..98T}. Such a scenario could also reproduce the profile of an individual double-peaked event. However, in both the one-zone and two-zone radiative interpretations, the two enhancement episodes or the activation times of the two emitting regions have to be prescribed. Therefore, these models do not by themselves predict the recurrent $\sim11.8$ d intra-subflare peak separation, nor do they explain the $\sim67$ d neighbouring-subflare spacing.

Magnetic reconnection has been invoked to explain rapid blazar flares and spectral changes in some sources \citep[e.g.,][]{2020NatCo..11.4176S,2023MNRAS.521L..53A}. Since the double-peaked structures in PKS 1424--418 are morphologically similar to the peak-in-peak variability reported in 3C 279 \citep{2020NatCo..11.4176S}, we examine whether the specific reconnection-driven jet-in-jet model can account for the observed day-scale intra-subflare peak separation. In this model, if the rise time of the slowly varying flare envelope is used to estimate the size of the reconnecting region, the comoving length of the reconnection layer can be estimated as
\begin{equation}
	l'
	\simeq
	\frac{\Gamma_j T_r \epsilon c}{1+z},
	\label{eq:lrec_simple}
\end{equation}
where $\Gamma_j \simeq \delta \sim 24$ is the bulk Lorentz factor of the jet, $T_r=10.05 \pm2.73$ d is the P3 rise time, $\epsilon \simeq 0.1$ is the dimensionless reconnection rate, and the redshift is $z=1.522$. We obtain $l'\simeq(2.5 \pm0.7)\times10^{16} {\rm cm}$. For a plasmoid, the duration of an individual plasmoid flare and the recurrence interval of large plasmoids can be approximated as $t_d\simeq\frac{(1+z)f l'}{\delta c}$, $\Delta t\simeq2.5\frac{(1+z)l'}{\delta c}$, respectively. Here $f\simeq0.1$ is the fractional size of the plasmoid relative to the reconnecting layer. The corresponding time-scales are therefore $t_d\simeq0.10 \pm0.03$ d and $\Delta t \simeq2.5 \pm0.7$ d. The predicted recurrence interval is therefore much shorter than the observed $\sim11.8$ d intra-subflare peak separation. If the plasmoid has additional Doppler boosting, as expected in the jet-in-jet, both time-scales would become even shorter. Therefore, the observed intra-subflare peak separation in PKS 1424--418 is unlikely to be explained by the monster-plasmoid recurrence time in the specific magnetic-reconnection model proposed for 3C 279. This does not rule out magnetic reconnection in a broader sense. In more general reconnection scenarios, large or slowly moving plasmoids, or the cumulative emission from an extended reconnection layer, may still produce longer-time-scale variability in blazar jets \citep[e.g.,][]{2019MNRAS.482...65C}.
However, such reconnection activity does not by itself provide a stable characteristic time-scale for the repeated $\sim11.8$ d intra-subflare peak separations or for the $\sim67 \pm 10$ d spacing between neighbouring sub-flares \citep[e.g.,][]{2007PhPl...14j0703L}.

In summary, the one-zone, two-zone, and magnetic-reconnection scenarios can each account for some aspects of an individual double-peaked event. However, in these interpretations, the temporal evolution of particle injection, emission-region activation, or local dissipation must generally be prescribed rather than predicted consistently. They therefore do not by themselves explain why similar double-peaked profiles recur in multiple sub-flares, nor do they naturally account for the $\sim67$ d spacing between neighbouring sub-flares. Repeated disturbances launched from the central engine could provide a possible external timing source, but there is currently no independent evidence for a corresponding stable or quasi-stable period. Moreover, such a picture would still require an additional mechanism to explain the similar $\sim11.8$ d intra-subflare peak separations.

Motivated by the limitations of the simple radiative and magnetic-reconnection interpretations discussed above, we further consider whether quasi-stationary structures in the jet may provide a possible physical mechanism for the observed recurrent timing pattern. Existing multiwavelength correlation studies provide useful constraints on this possibility. First, \citet{2021MNRAS.501.2504A} performed a correlation analysis of the multiwavelength light curves of PKS 1424--418 during MJD 56000--56600 and found that the $\gamma$-ray, optical, and infrared bands are strongly correlated with a near-zero time lag. This indicates that the enhancements of the low-energy synchrotron emission and the high-energy inverse-Compton emission occur nearly simultaneously, suggesting that the broadband flare activity is likely driven by a common jet disturbance or by dynamically connected emitting regions. Second, \citet{2024A&A...692A.203K} analysed the long-term Atacama Large Millimeter/submillimeter Array (ALMA) 90--350 GHz and \textit{Fermi}-LAT 0.1--200 GeV light curves and reported a significant mm/sub-mm--$\gamma$-ray correlation with no significant time delay. This further suggests a close association between the $\gamma$-ray activity and the compact mm/sub-mm core region, possibly indicating that the $\gamma$-ray emission occurs near this region. Since compact mm/sub-mm core region may be associated with standing or recollimation shocks in relativistic jets \citep{1997MNRAS.288..833K,2008Natur.452..966M}, these two classes of multiwavelength correlations jointly motivate a scenario in which the relevant disturbance operates near the  compact mm/sub-mm core region and may interact with quasi-stationary recollimation shocks. Two-dimensional special-relativistic magnetohydrodynamic (SRMHD) simulations by \citet{2015ApJ...809...38M} demonstrated that an over-pressured jet can develop a quasi-stationary multi-shock recollimation structure. In their simulations, the first recollimation shock is located at $13\text{--}15\,R_j$ (where $R_j$ denotes the jet radius), and both the background density and gas pressure reach local maxima there. In the downstream region of the shock ($15\text{--}17\,R_j$), the rest-mass density and gas pressure begin to decline. They drop by roughly a factor of five, reaching their minimum values and forming a pronounced rarefied zone. Subsequently, a second recollimation shock forms at $\sim 25 ~R_j$, exhibiting a structure similar to that of the first one. Their results suggest a quasi-periodic shock/rarefaction pattern, with a characteristic spacing of $\simeq 12 R_j$.
	
This shock/rarefaction pattern provides a possible framework for the observed recurrent sub-flares, in which a single disturbance propagates downstream and successively interacts with quasi-stationary recollimation shocks. Each shock disturbance interaction could enhance particle acceleration and give rise to a sequence of sub-flares \citep[e.g.,][]{2008Natur.452..966M}. When the disturbance first crosses the recollimation shock located at $13\text{--}15\,R_j$, it may produce the first peak (B1). The disturbance also excites a fast magnetosonic component at the shock, which propagates further downstream and enters the rarefied region at $15\text{--}17\,R_j$. The strong gradient in magnetosonic impedance in this region causes partial reflection of the wave \citep[e.g.,][]{2018ApJ...860..138P,2025arXiv251001742B}. The reflected component subsequently travels upstream and re-impinges on the same recollimation shock, potentially giving rise to the second peak (B2). Therefore, a single disturbance propagating through a multi-shock structure can provide a possible explanation for both the sequence of sub-flares and the double-peaked morphology within each sub-flare. This interpretation is not unique.

\begin{table*}
	\centering
	\caption{
		PKS 1424--418: parameters of the two major active epochs and their sub-flares.
		Column (2) presents the candidate interval, Columns (3)--(4) present the rise and decay time-scales,
		Columns (5)--(6) show their significances, and Column (7) presents the
		$\gamma$-ray peak energy flux.
	}
	\begin{tabular}{lcccccc}
		\toprule
		Sub-flare & MJD interval & $T_r$ (d) & $T_d$ (d)
		& $S_r$ & $S_d$
		& $\nu F_{\nu}$ (erg\,cm$^{-2}$\,s$^{-1}$) \\
		
		(1) & (2) & (3) & (4) & (5) & (6) & (7) \\
		\midrule
		P1 & 56124--56160 & $10.91\pm5.21$ & $8.53\pm2.98$
		& 3.06 & 4.23
		& $1.37\times10^{-10}$ \\
		P2 & 56205--56253 & $8.94\pm4.08$ & $8.94\pm4.01$
		& 3.35 & 3.10
		& $1.90\times10^{-10}$ \\
		P3 & 56384--56431 & $10.05\pm2.73$ & $13.95\pm4.75$
		& 5.33 & 4.00
		& $2.98\times10^{-10}$ \\
		\addlinespace
		P4 & 59764--59834 & $4.92\pm0.50$ & $4.28\pm0.87$
		& 16.90 & 6.72
		& $4.73\times10^{-10}$ \\
		P5 & 59834--59913 & $4.59\pm1.11$ & $6.23\pm1.40$
		& 7.00 & 6.00
		& $3.91\times10^{-10}$ \\
		P6 & 59913--59976 & $6.20\pm0.70$ & $7.47\pm1.28$
		& 14.80 & 7.16
		& $5.29\times10^{-10}$ \\
		\bottomrule
	\end{tabular}
	\label{tab3}
	
\end{table*}
	
Following the core-shift relation $r_{\rm core}\propto\nu^{-1}$, where $\nu$ is the observing frequency \citep{1998A&A...330...79L}, and given that the $\gamma$-ray emission in PKS 1424--418 is nearly cospatial with the (sub)mm core region \citep{2024A&A...692A.203K}, we adopt $r_{\mathrm{core}}\sim0.5$--$1~\mathrm{pc}$ as a core distance. This location is inside the adopted dusty torus radius $R_{\rm DT} \simeq2.6$ pc, and is therefore consistent with the EC-DT environment assumed in the SED modelling. For a bulk Lorentz factor of $\Gamma\sim\delta\simeq25$, the corresponding jet opening angle is $\theta_j\simeq1/\Gamma\simeq0.04~\mathrm{rad}$.
\begin{equation}
	R_j \simeq r_{\mathrm{core}} \, \theta_j \sim (6.17\text{--}12.34) \times 10^{16}~\mathrm{cm}.
\end{equation}
This suggests that the effective axial separation associated with the reflection of the fast magnetosonic component in the rarefied region is 
$L'_{\mathrm{dp}}\sim(1\text{--}2 R_{j})\simeq (6.17\text{--}24.68)\times10^{16}\,\mathrm{cm}$. The propagation speed of the disturbance in the comoving frame is assumed to be the fast magnetosonic speed $v'_{\mathrm{ms}}$ along the axis, which can be expressed as \citep{2013arXiv1301.5572S} (upper bound for $k \perp B$)
\begin{equation}
	\left(\frac{v'_{\mathrm{ms}}}{c}\right)^2 = \left(\frac{c_s}{c}\right)^2 + \left(\frac{v_A}{c}\right)^2 - \left(\frac{c_s}{c}\right)^2 \left(\frac{v_A}{c}\right)^2,
\end{equation}
where the sound speed is given by $c_s = v_{j}/M_{s}$, with $v_{j}$ the bulk jet velocity and $M_{s}$ the sound Mach number, and the Alfvén speed is determined by $(v_A/c)^2 = \sigma/(1+\sigma)$, where $\sigma$ denotes the jet magnetisation parameter. Adopting the jet core parameters from \citet{2015ApJ...809...38M}, namely $v_j \simeq 0.94 c$, $M_s \simeq 1.69$, and $\sigma \sim 10^{-2}$--$10^{-1}$, we obtain $v'_{\mathrm{ms}} \simeq 0.59 \pm 0.03 c$. The time interval between the double peaks can be calculated by
\begin{equation}
	\Delta t_{\mathrm{obs}} = \frac{1+z}{\delta}\,\frac{L'_{\mathrm{dp}}}{v'_{\mathrm{ms}}},
	\label{time}
\end{equation}
where $v'_{\mathrm{ms}}$ is the magnetosonic speed. Adopting $v'_{\mathrm{ms}} \simeq 0.59 \pm 0.03\,c$ and $L'_{\mathrm{dp}} \simeq (6.17\text{--}24.68)\times10^{16}~\mathrm{cm}$, the resulting interval between the two peaks is calculated to be approximately $10 \pm 6$ days. In addition, \citet{2015ApJ...809...38M} indicate a periodic spacing between consecutive recollimation shocks of approximately $12\,R_j$. Using this value as the effective distance $L' \simeq 12\,R_j$ in Equation (\ref{time}) yields an estimated neighbouring-subflare spacing of $\sim 74 \pm 24$ days.
	
The above geometry also implies that the two peaks within an individual double-peaked sub-flare are not produced at widely separated distances. The first peak is associated with the initial interaction between the propagating disturbance and a recollimation shock, whereas the second peak is produced when the reflected fast-magnetosonic component interacts again with the same shock. The two peaks should therefore experience nearly the same external photon field. For the DT component, the radial dependence of the photon energy density can be described as $U'_{\rm DT}(r)\propto [1+(r/R_{\rm DT})^4]^{-1}$, where $r$ is the emission-region distance  \citep{HayashidaApJ...754...114}. For $R_{\rm DT}\simeq2.6$ pc, the DT photon energy density varies by only a few per cent over the range $r=0.5-1$ pc. Even if neighbouring sub-flares are associated with adjacent recollimation shocks, the expected displacement is of order $12R_j\simeq0.24$--$0.48~{\rm pc}$, so the emission sites remain within the DT scale and no order-of-magnitude change in the DT seed photon density is expected.

These estimates are broadly consistent with the observed time-scales at the order-of-magnitude level. They should therefore be regarded as an approximate consistency check, rather than as a unique validation of the proposed scenario. As shown in Figure \ref{fig2} and Tables \ref{tab1} and \ref{tab3}, the $\gamma$-ray light curve of PKS 1424--418 during MJD 56117--56498 consists of multiple sub-flares (P1--P3), each showing a candidate double-peaked profile near its maximum. The observed intervals are broadly consistent with the estimated time-scales, with $67 \pm 10$ d versus $\sim74 \pm 24$ d for the neighbouring-subflare spacing and $11.8 \pm 2.9$ d versus $\sim10 \pm 6$ d for the intra-subflare peak separation.
	
\subsection{Variability comparison}

As shown in Figure \ref{fig2}, Epoch a (MJD 56117--56498) and Epoch b (MJD 59669--59978), which occurred nearly 3650 days apart, exhibit a striking morphological similarity. Because Epoch b lacks multiwavelength coverage, we do not attempt SED modelling for that epoch and instead restrict our analysis to a morphological comparison. Tables~\ref{tab1} and \ref{tab3} summarise the candidate intervals, intra-subflare peak separations, neighbouring-subflare spacings, rise and decay time-scales, flux-change significances, and peak $\gamma$-ray energy fluxes. These parameters are reported for six sub-flares: P1 (MJD 56124--56160), P2 (MJD 56205--56253), P3 (MJD 56384--56431), P4 (MJD 59764--59834), P5 (MJD 59834--59913), and P6 (MJD 59913--59976). We estimate the flux doubling time $\tau$ from adjacent flux measurements following \citet{2011A&A...530A..77F}
\begin{equation}
	F(t_2) \;=\; F(t_1)\, 2^{\frac{t_2 - t_1}{\tau}},
	\end{equation}
where $F(t_1)$ and $F(t_2)$ denote the fluxes at times $t_1$ and $t_2$, respectively. We use only pairs with statistical significance $>3$, defined by
\begin{equation}
	\mathrm{Sig} \;=\; \left| \frac{F(t_2)-F(t_1)}{\sigma_{F(t_1)}} \right|,
\end{equation}
where $\sigma_{F(t_1)}$ denotes the uncertainty of the flux measured at time $t_1$. Among these, the shortest doubling times define the rise and decay time-scales. When the flux increases exponentially with time, the rise time-scale follows from the doubling time via $T_r=\tau/\ln 2$; analogously, the decay time-scale is $T_d=\tau/\ln 2$.
	
In summary, the sub-flares of the two active epochs exhibit a high degree of similarity in both morphology and temporal characteristics: the rise and decay phases are nearly symmetric, the durations are comparable, and the intra-subflare peak separations remain stable ($11.8 \pm 2.9$ days). In addition, the spacings between neighbouring sub-flares within each active epoch are also similar, with an average interval of $67 \pm 10$ days. The ratios of $T_{r}$ to $T_{d}$ are generally close to unity, indicating that the flares are roughly symmetric and that the time-scales of particle acceleration and cooling are comparable. Such repetitive morphology may hint at a possible long-time-scale modulation \citep{2005ApJ...635L..17L, Wang2026}. Further active epochs are required to determine whether this similarity represents a recurrent long-term behaviour.
	
\section{Conclusions}\label{Conclusion}

In this work, we investigated a hint of recurrent double-peaked $\gamma$-ray sub-flares in PKS 1424--418 using multiwavelength light curves, detailed \textit{Fermi}-LAT timing analysis, and quasi-simultaneous broadband SED modelling. The SED modelling suggests that the flaring states are mainly characterised by enhanced Doppler factors, together with changes in the inferred energy densities and jet powers. The two peaks, B1 and B2, of the P3 candidate show similar radiative properties, while the comparably high Doppler factor in the comparison flaring state A suggests that Doppler boosting alone cannot explain the hint of double-peaked structure. The timing analysis identifies five fit-supported double-peaked candidates and one weaker double-peak, with a characteristic intra-subflare peak separation of $11.8 \pm2.9$ d and a neighbouring-subflare spacing of $67 \pm10$ d. The ACF, PACF, and DPS diagnostics show only weak local enhancements near these time-scales, and are therefore used only as auxiliary checks rather than as independent evidence for a unique time delay or strict periodicity. The morphology-based red-noise Monte Carlo tests further give false-alarm probabilities of $p_{\rm a}=0.0216$ and $p_{\rm b}=0.0006$ for the two active epochs, respectively. In the simulations covering the entire LAT light curve, no red-noise realisation reproduces the joint occurrence of both epoch-like structures. These results indicate that the hint of recurrent double-peaked structure is not easily reproduced by the adopted red-noise null model, although this should not be interpreted as an independent Gaussian-equivalent significance for each individual sub-flare. With only two major active epochs observed, further events are needed to determine whether similar activity patterns recur on comparable long time-scales.

Previous multiwavelength correlation studies have reported near-zero lags between the optical/infrared and $\gamma$-ray bands, as well as between the mm/sub-mm and $\gamma$-ray bands \citep{2021MNRAS.501.2504A,2024A&A...692A.203K}, suggesting that the recurrent activity may be associated with a common jet disturbance near the compact mm/sub-mm core region. Motivated by this and by simulations of over-pressured jets that form chains of quasi-stationary recollimation shocks \citep{2015ApJ...809...38M}, we discuss a possible, non-unique interpretation in which a disturbance propagates through a multi-shock jet structure. In this scenario, repeated interactions between the disturbance and the shock structure may qualitatively link the double-peaked profiles to the $\sim 11.8$ d time-scale and the neighbouring-subflare spacing to the $\sim 67$ d time-scale. Future multiwavelength monitoring, high-resolution Very Long Baseline Interferometry (VLBI) imaging, and time-dependent SRMHD simulations will be essential for testing the multi-shock interpretation and for constraining the internal structure of the $\gamma$-ray emission region.
	
\section*{Acknowledgements}
This work was supported by the National Natural Science Foundation of China under grants 12494572 and 12221003. We acknowledge the use of data, analysis tools, and services from the Open Universe platform, the High Energy Astrophysics Science Archive Research Center (HEASARC), the \textit{Fermi} Science Tools, the Astrophysics Data System (ADS), and the NASA/IPAC Extragalactic Database (NED). 
	
\section*{Data Availability}
	
The data underlying this article are available in public archives. The \textit{Fermi}-LAT data are available from the Fermi Science Support Center, the \textit{Swift} data are available from the HEASARC archive, and the SMARTS optical data are available from the SMARTS blazar monitoring programme website. Derived data products will be shared on reasonable request to the corresponding author.

\bsp
\label{lastpage}

\begin{thebibliography}{99}
	
\bibitem[Abdo et al.(2010a)]{2010ApJ...716...30A} Abdo A.~A., Ackermann M., Agudo I., et al.\ 2010a, \apj, 716, 30. doi: \href{https://doi.org/10.1088/0004-637X/716/1/30}{10.1088/0004-637X/716/1/30}

\bibitem[Abdo et al.(2010b)]{2010ApJ...722..520A} Abdo A. A., Ackermann M., Ajello M., et al.\ 2010b, \apj, 722, 520. doi: \href{https://doi.org/10.1088/0004-637X/722/1/520}{10.1088/0004-637X/722/1/520}

\bibitem[Abdollahi et al.(2020)]{2020ApJS..247...33A} Abdollahi S., Acero F., Ackermann M., et al.\ 2020, \apjs, 247, 33. doi: \href{https://doi.org/10.3847/1538-4365/ab6bcb}{10.3847/1538-4365/ab6bcb}

\bibitem[Abdollahi et al.(2023)]{2023ApJS..265..31} Abdollahi S., Ajello M., Baldini L., et al.\ 2023, \apjs, 265, 31. doi: \href{https://doi.org/10.3847/1538-4365/acbb6a}{10.3847/1538-4365/acbb6a}

\bibitem[Abhir et al.(2021)]{2021MNRAS.501.2504A} Abhir J., Joseph J., Patel S.~R., et al.\ 2021, \mnras, 501, 2504. doi: \href{https://doi.org/10.1093/mnras/staa3639}{10.1093/mnras/staa3639}

\bibitem[Agarwal et al.(2023)]{2023MNRAS.521L..53A} Agarwal S., Banerjee B., Shukla A., Roy J., Acharya S., Vaidya B., Chitnis V.~R., et al.\ 2023, MNRAS, 521, L53. doi: \href{https://doi.org/10.1093/mnrasl/slad023}{10.1093/mnrasl/slad023}

\bibitem[Agarwal et al.(2024)]{2024ApJ...968L...1A}
Agarwal S., Shukla A., Mannheim K., Vaidya B., Banerjee B.\ 2024, ApJL, 968, L1. doi: \href{https://doi.org/10.3847/2041-8213/ad4994}{10.3847/2041-8213/ad4994}

\bibitem[Agarwal et al.(2025)]{2025MNRAS.537..332A} Agarwal S., Shukla A., Sharma P.\ 2025, \mnras, 537, 332. doi: \href{https://doi.org/10.1093/mnras/staf021}{10.1093/mnras/staf021}

\bibitem[Akaike(1974)]{1974ITAC...19..716A}
Akaike H., 1974, IEEE Trans. Autom. Control, 19, 716. doi: \href{https://doi.org/10.1109/TAC.1974.1100705}{10.1109/TAC.1974.1100705}

\bibitem[Arnaud(1996)]{1996ASPC..101...17A} Arnaud K.~A.\ 1996, in ASP Conf.\ Ser.\ 101, Astronomical Data Analysis Software and Systems V, ed.\ G.~H.\ Jacoby \& J.\ Barnes (San Francisco, CA: ASP), 17

\bibitem[Atwood et al.(2009)]{2009ApJ...697.1071A} Atwood W.~B., Abdo A.~A., Ackermann M., et al.\ 2009, \apj, 697, 1071. doi: \href{https://doi.org/10.1088/0004-637X/697/2/1071}{10.1088/0004-637X/697/2/1071}

\bibitem[Barnacka et al.(2011)]{2011A&A...528L...3B} Barnacka A., Glicenstein J.-F., Moudden Y.\ 2011, \aap, 528, L3. doi: \href{https://doi.org/10.1051/0004-6361/201016175}{10.1051/0004-6361/201016175}

\bibitem[Barnacka et al.(2015)]{2015ApJ...809..100B} Barnacka A., Geller M. J., Dell'Antonio I. P., Benbow W.\ 2015, \apj, 809, 100. doi: \href{https://doi.org/10.1088/0004-637X/809/1/100}{10.1088/0004-637X/809/1/100}

\bibitem[Barthelmy et al.(2005)]{2005SSRv..120..143B} Barthelmy S.~D., Barbier L.~M., Cummings J.~R., et al.\ 2005, \ssr, 120, 143. doi: \href{https://doi.org/10.1007/s11214-005-5096-3}{10.1007/s11214-005-5096-3}

\bibitem[Bonning et al.(2012)]{2012ApJ...756...13B} Bonning E., Urry C.~M., Bailyn C., et al.\ 2012, \apj, 756, 13. doi: \href{https://doi.org/10.1088/0004-637X/756/1/13}{10.1088/0004-637X/756/1/13}

\bibitem[Boula et al.(2025)]{2025arXiv251001742B} Boula S., Tavecchio F., Bodo G., Vlahakis N., Coppi P., Costa A., Sciaccaluga A.\ 2025, arXiv:2510.01742. doi: \href{https://doi.org/10.48550/arXiv.2510.01742}{arXiv.2510.01742}

\bibitem[B{\"o}ttcher \& Chiang(2002)]{2002ApJ...581..127B} B{\"o}ttcher M., Chiang J.\ 2002, \apj, 581, 127. doi: \href{https://doi.org/10.1086/344155}{10.1086/344155}

\bibitem[Brockwell \& Davis(2002)]{2002itsa.book.....B} Brockwell P. J., Davis R. A.\ 2002, Introduction to Time Series and Forecasting, 2nd edn. New York: Springer. doi: \href{https://doi.org/10.1007/b97391}{10.1007/b97391}

\bibitem[Brown(2013)]{2013MNRAS.431..824B} Brown A.~M.\ 2013, \mnras, 431, 824. doi: \href{https://doi.org/10.1093/mnras/stt218}{10.1093/mnras/stt218}

\bibitem[Burnham \& Anderson(2002)]{2002msma.book.....B} Burnham K.~P., Anderson D.~R., 2002, Model Selection and Multimodel Inference: A Practical Information-Theoretic Approach, 2nd edn. Springer, New York. doi: \href{https://doi.org/10.1007/b97636}{10.1007/b97636}

\bibitem[Burrows et al.(2005)]{2005SSRv..120..165B} Burrows D.~N., Hill J.~E., Nousek J.~A., et al.\ 2005, \ssr, 120, 165. doi: \href{https://doi.org/10.1007/s11214-005-5097-2}{10.1007/s11214-005-5097-2}

\bibitem[Buson et al.(2014)]{2014A&A...569A..40B} Buson S., Longo F., Larsson S., et al.\ 2014, \aap, 569, A40. doi: \href{https://doi.org/10.1051/0004-6361/201423367}{10.1051/0004-6361/201423367}

\bibitem[Christie et al.(2019)]{2019MNRAS.482...65C} Christie I.~M., Petropoulou M., Sironi L., Giannios D.\ 2019, \mnras, 482, 65. doi: \href{https://doi.org/10.1093/mnras/sty2636}{10.1093/mnras/sty2636}

\bibitem[Connolly(2015)]{2015arXiv150306676C} Connolly S.~D.\ 2015, arXiv e-prints, arXiv:1503.06676. doi: \href{https://doi.org/10.48550/arXiv.1503.06676}{10.48550/arXiv.1503.06676}

\bibitem[Eisenstein \& Hut(1998)]{1998ApJ...498..137E} Eisenstein D.~J., Hut P., 1998, ApJ, 498, 137. doi: \href{https://doi.org/10.1086/305535}{10.1086/305535}

\bibitem[Emmanoulopoulos et al.(2013)]{2013MNRAS.433..907E} Emmanoulopoulos D., McHardy I. M., Papadakis I. E.\ 2013, \mnras, 433, 907. doi: \href{https://doi.org/10.1093/mnras/stt764}{10.1093/mnras/stt764}

\bibitem[Fan \& Cao(2004)]{2004ApJ...602..103F} Fan Z.~H., Cao X.\ 2004, \apj, 602, 103. doi: \href{https://doi.org/10.1086/380902}{10.1086/380902}

\bibitem[Foschini et al.(2011)]{2011A&A...530A..77F} Foschini L., Ghisellini G., Tavecchio F., Bonnoli G., Stamerra A.\ 2011, \aap, 530, A77. doi: \href{https://doi.org/10.1051/0004-6361/201117064}{10.1051/0004-6361/201117064}

\bibitem[Fossati et al.(1998)]{1998MNRAS.299..433F} Fossati G., Maraschi L., Celotti A., et al.\ 1998, \mnras, 299, 433. doi: \href{https://doi.org/10.1046/j.1365-8711.1998.01828.x}{10.1046/j.1365-8711.1998.01828.x}

\bibitem[Gehrels et al.(2004)]{2004ApJ...611.1005G} Gehrels N., Chincarini G., Giommi P.~E., et al.\ 2004, \apj, 611, 1005. doi: \href{https://doi.org/10.1086/422091}{10.1086/422091}

\bibitem[Ghisellini et al.(2005)]{2005A&A...432..401G} Ghisellini G., Tavecchio F., Chiaberge M.\ 2005, \aap, 432, 401. doi: \href{https://doi.org/10.1051/0004-6361:20041404}{10.1051/0004-6361:20041404}

\bibitem[Ghisellini et al.(2014)]{2014Natur.515..376G} Ghisellini G., Tavecchio F., Maraschi L., Celotti A., Sbarrato T.\ 2014, \nat, 515, 376. doi: \href{https://doi.org/10.1038/nature13856}{10.1038/nature13856}

\bibitem[Hayashida et al.(2012)]{HayashidaApJ...754...114} Hayashida M., Madejski G.~M., Nalewajko K., et al.\ 2012, ApJ, 754, 114. doi: \href{https://doi.org/10.1088/0004-637X/754/2/114}{10.1088/0004-637X/754/2/114}

\bibitem[Jones et al.(1974)]{1974ApJ...188..353J} Jones T.~W., O'Dell S.~L., Stein W.~A.\ 1974, \apj, 188, 353. doi: \href{https://doi.org/10.1086/152724}{10.1086/152724}

\bibitem[Kass \& Raftery(1995)]{1995JASA...90..773K} Kass R.~E., Raftery A.~E., 1995, J. Am. Stat. Assoc., 90, 773. doi: \href{https://doi.org/10.1080/01621459.1995.10476572}{10.1080/01621459.1995.10476572}

\bibitem[Kim et al.(2024)]{2024A&A...692A.203K} Kim D.-W., Ros E., Kadler M., Krichbaum T.~P., Zhao G.-Y., R{\"o}sch F., et al.\ 2024, \aap, 692, A203. doi: \href{https://doi.org/10.1051/0004-6361/202451773}{10.1051/0004-6361/202451773}

\bibitem[Komissarov \& Falle(1997)]{1997MNRAS.288..833K} Komissarov S.~S., Falle S.~A.~E.~G.\ 1997, \mnras, 288, 833. doi: \href{https://doi.org/10.1093/mnras/288.4.833}{10.1093/mnras/288.4.833}

\bibitem[Liodakis et al.(2020)]{2020ApJ...902...61L} Liodakis I., Blinov D., Jorstad S.~G., Arkharov A.~A., Di~Paola A., Efimova N.~V., et al.\ 2020, \apj, 902, 61. doi: \href{https://doi.org/10.3847/1538-4357/abb1b8}{10.3847/1538-4357/abb1b8}

\bibitem[Lobanov(1998)]{1998A&A...330...79L} Lobanov A.~P.\ 1998, \aap, 330, 79. doi: \href{https://doi.org/10.48550/arXiv.astro-ph/9712132}{10.48550/arXiv.astro-ph/9712132}

\bibitem[Loureiro et al.(2007)]{2007PhPl...14j0703L} Loureiro N.~F., Schekochihin A.~A., Cowley S.~C.\ 2007, Physics of Plasmas, 14, 100703. doi: \href{https://doi.org/10.1063/1.2783986}{10.1063/1.2783986}

\bibitem[Lu \& Zhou(2005)]{2005ApJ...635L..17L} Lu J.-F., Zhou B.-Y.\ 2005, \apjl, 635, L17. doi: \href{https://doi.org/10.1086/499333}{10.1086/499333}

\bibitem[Marscher et al.(2008)]{2008Natur.452..966M} Marscher A.~P., Jorstad S.~G., D'Arcangelo F.~D., Smith P.~S., Williams G.~G., Larionov V.~M., et al.\ 2008, \nat, 452, 966. doi: \href{https://doi.org/10.1038/nature06895}{10.1038/nature06895}

\bibitem[Mastichiadis \& Kirk(1997)]{1997A&A...320...19M} Mastichiadis A., Kirk J.~G.\ 1997, A\&A, 320, 19. doi: \href{https://doi.org/10.48550/arXiv.astro-ph/9610058}{10.48550/arXiv.astro-ph/9610058}

\bibitem[Max-Moerbeck et al.(2014)]{2014MNRAS.445..428M} Max-Moerbeck W., Hovatta T., Richards J. L., et al.\ 2014, \mnras, 445, 428. doi: \href{https://doi.org/10.1093/mnras/stu1749}{10.1093/mnras/stu1749}

\bibitem[Mizuno et al.(2015)]{2015ApJ...809...38M} Mizuno Y., G{\'o}mez J.~L., Nishikawa K.-I., Meli A., Hardee P.~E., Rezzolla L.\ 2015, \apj, 809, 38. doi: \href{https://doi.org/10.1088/0004-637X/809/1/38}{10.1088/0004-637X/809/1/38}

\bibitem[Nakagawa \& Mori(2013)]{2013ApJ...773..177N} Nakagawa K., Mori M.\ 2013, \apj, 773, 177. doi: \href{https://doi.org/10.1088/0004-637X/773/2/177}{10.1088/0004-637X/773/2/177}

\bibitem[Padovani \& Giommi(1995)]{1995ApJ...444..567P} Padovani P., Giommi P.\ 1995, \apj, 444, 567. doi: \href{https://doi.org/10.1086/175631}{10.1086/175631}

\bibitem[Prince et al.(2019)]{2019ApJ...883..137P} Prince R., Gupta N., Nalewajko K.\ 2019, \apj, 883, 137. doi: \href{https://doi.org/10.3847/1538-4357/ab3afa}{10.3847/1538-4357/ab3afa}

\bibitem[Provornikova et al.(2018)]{2018ApJ...860..138P} Provornikova E., Laming J.~M., Lukin V.~S.\ 2018, \apj, 860, 138. doi: \href{https://doi.org/10.3847/1538-4357/aac1c1}{10.3847/1538-4357/aac1c1}

\bibitem[Raiteri et al.(2015)]{2015MNRAS.454..353R} Raiteri C.~M., Stamerra A., Villata M., et al.\ 2015, \mnras, 454, 353. doi: \href{https://doi.org/10.1093/mnras/stv1884}{10.1093/mnras/stv1884}

\bibitem[Raiteri et al.(2017)]{2017Natur.552..374R} Raiteri C.~M., Villata M., Acosta-Pulido J.~A., Agudo I., Arkharov A.~A., Bachev R., Baida G.~V., et al.\ 2017, \nat, 552, 374. doi: \href{https://doi.org/10.1038/nature24623}{10.1038/nature24623}

\bibitem[Raiteri et al.(2021)]{2021MNRAS.501.1100R}
Raiteri C.~M., Villata M., Carosati D., Ben{\'i}tez E., Kurtanidze S.~O., Gupta A.~C., Mirzaqulov D.~O., et al.\ 2021, \mnras, 501, 1100. doi: \href{https://doi.org/10.1093/mnras/staa3561}{10.1093/mnras/staa3561}

\bibitem[Roming et al.(2005)]{2005SSRv..120...95R} Roming P.~W.~A., Kennedy T.~E., Mason K.~O., et al.\ 2005, \ssr, 120, 95. doi: \href{https://doi.org/10.1007/s11214-005-5095-4}{10.1007/s11214-005-5095-4}

\bibitem[Scargle et al.(2013)]{2013ApJ...764..167S} Scargle J.~D., Norris J.~P., Jackson B., Chiang J., 2013, ApJ, 764, 167. doi: \href{https://doi.org/10.1088/0004-637X/764/2/167}{10.1088/0004-637X/764/2/167}

\bibitem[Schwarz(1978)]{1978AnSta...6..461S} Schwarz G.\ 1978, Ann. Statist., 6, 461. doi: \href{https://doi.org/10.1214/aos/1176344136}{10.1214/aos/1176344136}

\bibitem[Shukla \& Mannheim(2020)]{2020NatCo..11.4176S} Shukla A., Mannheim K.\ 2020, Nature Communications, 11, 4176. doi: \href{https://doi.org/10.1038/s41467-020-17912-z}{10.1038/s41467-020-17912-z}

\bibitem[Spruit(2013)]{2013arXiv1301.5572S} Spruit H.~C.\ 2013, arXiv:1301.5572. doi: \href{https://doi.org/10.48550/arXiv.1301.5572}{10.48550/arXiv.1301.5572}


\bibitem[Tavecchio \& Ghisellini(2008)]{2008MNRAS.385L..98T} Tavecchio F., Ghisellini G.\ 2008, \mnras, 385, L98. doi: \href{https://doi.org/10.1111/j.1745-3933.2008.00441.x}{10.1111/j.1745-3933.2008.00441.x}

\bibitem[Tramacere et al.(2009)]{2009A&A...501..879T} Tramacere A., Giommi P., Perri M., et al.\ 2009, \aap, 501, 879. doi: \href{https://doi.org/10.1051/0004-6361/200810865}{10.1051/0004-6361/200810865}

\bibitem[Tramacere(2020)]{2020ascl.soft09001T} Tramacere A.\ 2020, \textit{Astrophysics Source Code Library}, ascl:2009.001

\bibitem[Tramacere et al.(2011)]{2011ApJ...739...66T} Tramacere A., Massaro E., Taylor A.\ M.\ 2011, \apj, 739, 66. doi: \href{https://doi.org/10.1088/0004-637X/739/2/66}{10.1088/0004-637X/739/2/66}

\bibitem[Urry \& Padovani(1995)]{1995PASP..107..803U} Urry C.~M., Padovani P.\ 1995, \pasp, 107, 803. doi: \href{https://doi.org/10.1086/133630}{10.1086/133630}

\bibitem[Uttley et al.(2002)]{2002MNRAS.332..231U} Uttley P., McHardy I.~M., Papadakis I.~E.\ 2002, MNRAS, 332, 231. doi: \href{https://doi.org/10.1046/j.1365-8711.2002.05298.x}{10.1046/j.1365-8711.2002.05298.x}

\bibitem[Wagner et al.(2021)]{Wagner2021} Wagner S.\ M., Burd P.\ R., Dorner D., et al.\ 2021, PoS, ICRC2021, 868. doi: \href{https://doi.org/10.22323/1.395.0868}{10.22323/1.395.0868}

\bibitem[Wang et al.(2026)]{Wang2026} Wang N., Ren G., Zhang S., et al.\ 2026, J.\ High Energy Astrophys., 51, 100529. doi: \href{https://doi.org/10.1016/j.jheap.2025.100529}{10.1016/j.jheap.2025.100529}

\bibitem[Wood et al.(2017)]{2017ICRC...301W} Wood M., et al.\ 2017, in Proc.\ 35th Int.\ Cosmic Ray Conf.\ (ICRC2017), PoS(ICRC2017)824. doi: \href{https://doi.org/10.22323/1.301.0824}{10.22323/1.301.0824}
\end{thebibliography}
\end{document}